\documentclass[11pt,a4paper]{article}
\usepackage[centertags]{amsmath}
\usepackage{amsfonts}
\usepackage{amssymb}
\usepackage{amsthm}
\usepackage{epsfig}
\usepackage{ae} % Umlauts

\usepackage[usenames]{color}

\theoremstyle{plain}
\newtheorem{thm}{Theorem}[section]
\newtheorem{lem}[thm]{Lemma}
\newtheorem{cor}[thm]{Corollary}
\newtheorem{prop}[thm]{Proposition}
\theoremstyle{definition}
\newtheorem{defi}[thm]{Definition}

\theoremstyle{remark}
\newtheorem{exmp}[thm]{Example}
\newtheorem{rem}[thm]{Remark}
\newcommand{\DR}{\mathbb{R}}
\newcommand{\R}{\mathbb{R}}
\newcommand{\No}{\mathbb{N}_0}
\newcommand{\N}{\mathbb{N}}

\DeclareMathOperator{\supp}{supp}

\begin{document}

\title{Model-Free Finance and Non-Lattice Integration}
\author{Christian Bender$^1$, Sebastian Ferrando$^2$, and Alfredo Gonzalez$^3$}
\maketitle
\footnotetext[1]{Saarland University, Department of Mathematics,
Postfach 151150, D-66041 Saarbr\"ucken, Germany, {\tt
bender@math.uni-sb.de}} \footnotetext[2]{Ryerson University, Department of Mathematics, 350 Victoria Street, Toronto, Ontario M5B 2K3, Canada {\tt ferrando@ryerson.ca}}
\footnotetext[3]{Universidad Nacional de Mar del Plata, Department of Mathematics,  Funes 3350, Mar del
Plata 7600, Argentina {\tt algonzal@mdp.edu.ar}}

\begin{abstract}
Starting solely with a set of possible prices for a traded asset $S$ (in infinite discrete time) expressed in units of a numeraire, we explain how to construct a Daniell type of integral representing prices of integrable functions depending on the asset. Such functions include the values of simple dynamic portfolios obtained by trading with $S$ and the numeraire. The space of elementary integrable functions, i.e. the said portfolio values, is not a vector lattice. It then follows that the integral is not classical, i.e. it is not associated to a measure. The essential ingredient in constructing the integral is a weak version of the no-arbitrage condition but here expressed in terms of properties of the trajectory space. We also discuss the continuity conditions imposed by Leinert (Archiv der Mathematik, 1982) and K\"onig (Mathematische Annalen, 1982) in the abstract theory of non-lattice integration from a 
financial point of view and establish some connections between these continuity conditions and the existence of martingale measures.  \\[0.1cm]
 \\[0.1cm]
\end{abstract}

\section{Introduction}

By no-arbitrage considerations, the fundamental theorem of asset pricing (FTAP) in its classical form \cite{dalang, delbaen}  restricts 
 stock price models in the option pricing framework to stochastic processes which can be transformed into 
(local) martingales via an equivalent change of the probability measure. Partially motivated by the aim to incorporate fractional Brownian motion and related non-semimartingale processes into financial modelling, no-arbitrage results and option pricing results have been derived  beyond the framework of the  classical FTAP. Results in this direction can be obtained by adding market frictions such as 
proportional transaction costs, e.g.
\cite{guasoni,guasoni2, czichowsky}, or by restricting the class of admissible trading  strategies, e.g. \cite{schoenmakers, bender2, cheridito, jarrow, bender1}. In a nutshell, no-arbitrage results in these generalized frameworks do typically 
require some conditions on the long-time behaviour of the paths. These include the stickiness condition \cite{guasoni2} or the conditional full support  property \cite{guasoni}. Additionally, conditions 
on the fine structure of the paths may be required to prevent arbitrage opportunities by trading on arbitrarily small time intervals including conditions on the pathwise quadratic variation \cite{schoenmakers, bender1} or the two-way crossing property \cite{bender2,peyre}. While in these works a first shift of paradigms from  the probabilistic martingale condition to paths properties may be visible, the paths properties are still formulated with respect to a reference probability measure.

Another stream of research completely dispenses with any a priori given probability measure. Initiated by the work of Vovk, see e.g. 
\cite{shafer, vovk1, vovk}, an outer measure is introduced via superhedging with (infinitely many) simple portfolios with positive terminal wealth.  
Vovk's outer measure and variants thereof have subsequently been applied by several authors in order to define pathwise stochastic integrals \cite{vovk,perkowski} for trading in continuous time or to study the duality between (non-dominated) martingale measures and superhedging prices in continuous time \cite{beiglbock, bartl, bartl2}. In finite discrete time, connections between martingale measures, arbitrage, and pricing in a model-free setting have been established e.g. in 
\cite{coxHouObloj, acciaio, burzoni, burzoni2, burzoni3, ferrando2}. An important observation by Burzoni et al. \cite{burzoni, burzoni2, burzoni3} is that one has to move to a reduced trajectory set  in finite discrete time  in order to come up with   general model-free versions of the FTAP and of the superhedging theorem. This reduction can be achieved via the notion of an arbitrage aggregator. The null sets of Vovk's outer measure in continuous time and the trajectories charged by the arbitrage aggregator in finite discrete time can be viewed as a mechanism to identify negligible sets by  arbitrage considerations  which replace the null sets of the reference probability measure in the classical probabilistic setting. We 
also refer to Cassese \cite{cassese} for implications on arbitrage and pricing starting from abstract collections of negligible sets. 

In this paper, we put ourselves in the context of trading in infinite discrete time. Given any trajectory set, which satisfies a trajectory-wise version of the law of one price, we interpret the initial endowment of a finite linear combination of buy-and-hold strategies as  elementary integral of the portfolio value at maturity. The corresponding space of elementary integrable functions (the said portfolios values) fails to have the lattice property (except in binary models). Thus, any extension of this elementary integral can be viewed as a non-classical integral and need not be related to integration with respect to a probability measure. For the extension step we make us of the abstract theory of non-lattice integration developed in the 1980's by Leinert \cite{leinert} and K\"onig \cite{konig}, which to the best of our knowledge has not yet been 
applied in model-free finance with the exception of \cite{ferrando}. The construction of Leinert and K\"onig relies on two key operators which can be interpreted as superhedging functionals in the financial context. One of them plays the role of a (semi-)norm which is applied for a continuous extension in Leinert's approach. The other one is used for the definition of outer and inner integrals in K\"onig's approach. The norm operator becomes the analogue of Vovk's outer measure in our infinite discrete time model-free approach, if restricted to indicator functions. 

We show that Leinert's continuity condition, which is required for the continuous extension step, is equivalent to a concept of absence of strict model-independent arbitrage (cp. \cite{acciaio, davis}) in a suitable class of generalized strategies. This concept of absence of arbitrage may be considered as a minimal condition for a `reasonable' trajectory set in our context. The countably subadditive norm operator also induces a notion of null sets in a canonical way. We illustrate by an example that Leinert's continuity condition is not sufficient to guarantee absence of arbitrage with buy-and-hold strategies if these null sets are considered as negligible. Mathematically, this type of arbitrage can arise because the norm operator may fail to coincide with the (outer) integral on the set of positive integrable functions, indicating that superhedging operators defined in the spirit of Vovk's outer measure may be inadequate for pricing purposes in such situations. We explain that absence of arbitrage with respect to the null sets of the norm operator can be obtained under K\"onig's stronger continuity condition on the elementary integral, which has been studied in detail in the abstract setting of non-lattice integration in \cite{konig}. Under K\"onig's condition, the extended integral can also be interpreted as a linear, strictly positive pricing system on a suitable subspace of all financial positions. Compare e.g. \cite{biagini, kassberger, riedel}
for axiomatic approaches to pricing systems, but note that their results do not apply in our setting which lacks the lattice property. The development of conditional non-lattice integrals that may substitute conditional expectations for pricing purposes in our framework is detailed in the companion paper \cite{bender}. 

The paper is organized as follows: In Section \ref{sec:setting}, we introduce 
the tra\-jectory-based financial market model. In Sections \ref{sec:leinert} 
and \ref{sec:konig}, we review Leinert's and K\"onig's constructions of non-lattice integration in our financial context. In particular, we point to the importance of K\"onig's continuity condition when considering the `superhedging null sets' of the norm operator  as negligible sets for pricing purposes and for absence of arbitrage. The remaining sections are devoted to the derivation of sufficient conditions for K\"onig's continuity condition.
In Section \ref{sec:L_vs_K}, we prove that a conditional version of Leinert's condition (when restarting the model at any later time) implies the conditional version of K\"onig's condition, although such an implication is not true for the unconditional counterparts. Finally, in Section \ref{sec:martingale}, we establish some connections to the existence of martingale measures. We show that the existence of a  martingale measure with a tree-like support implies Leinert's condition, while neither does the converse hold nor does the existence of a martingale measure with a tree-like support imply K\"onig's condition. Nonetheless, we can apply martingale arguments in conjunction with the results of Section \ref{sec:L_vs_K} to prove an easy-to-check sufficient criterion for K\"onig's condition and, thus, for a satisfactory arbitrage theory with respect to the collection of null sets which is intrinsically connected to the trajectory set by the superhedging norm: It requires that trajectories cannot  strictly increase (or, decrease) in all possible scenarios for the next time step and, additionally, a simple (topological) completeness property of the trajectory set.

\section{The financial market}\label{sec:setting}

For simplicity, we consider a financial market in discounted units, in which one can trade in a constant bond and one stock in discrete, but infinite time.
Given an initial price $s_0\in \R$ of the stock, the conceivable evolutions
of the stock price over time are collected in a set of trajectories
$
\mathcal{S}
$
which can be any subset of the space of real valued sequences $(S_j)_{j\in \No}$ such that $S_0=s_0$.

As usual, a \emph{simple portfolio} consists of an initial endowment $V\in \R$, a maturity $n\in \N$, and a finite sequence $H$ of   functions 
$H_i:\mathcal{S}\rightarrow \mathbb{R}$, $0\leq i \leq n-1$ which models the numbers of shares held by the investor in the stock between time $i$ and $i+1$. We assume throughout the paper, that simple portfolios are \emph{nonanticipating} in the sense that $H_i$ can be represented in the form 
$$
H_i(S)=h_i(S_0,\ldots, S_i),\quad S\in \mathcal{S},
$$
for some function $h_i:\mathbb{R}^{i+1}\rightarrow \mathbb{R}$. We emphasize that we do not assume that the functions $h_i$ are Borel-measurable.  So our class of simple portfolios is larger than the one usually considered in probabilistic finance,  and we refer to \cite{burzoni3} for the discussion of some subtle measurability issues in pointwise arbitrage theory. We note that various portfolio restrictions can be incorporated into the theory of non-lattice integration for model-free finance as detailed in \cite{bender}, but in the present paper we consider all nonanticipating portfolios as admissible.
The self-financing condition entails that the \emph{wealth
process} of a simple portfolio is given by
$$
\Pi^{V,n,H}_j(S)= V+\sum_{i=0}^{\min\{j,n\}-1} H_i(S)(S_{i+1}-S_i),\quad j\in \No, \quad S\in \mathcal{S}.
$$
Note that the limit $\Pi^{V,n,H}_\infty:=\lim_{j\rightarrow \infty} \Pi^{V,n,H}_j=\Pi^{V,n,H}_n$ always exists.
We call a simple portfolio \emph{positive}, if $V\geq 0$ and if for every $S\in \mathcal{S}$,
$$
\Pi^{V,n,H}_\infty(S)\geq 0.
$$

A \emph{generalized portfolio} consists of a sequence $(V_m, n_m, H_m)_{m\in \No}$ of simple portfolios, such that $(V_m, n_m, H_m)$ is positive for every $m\geq 1$. If additionally
$\Pi^{V_0,n_0,H_0}_j \equiv 0 $ for every $j\in \No$, we speak of a \emph{positive generalized portfolio} and  sometimes write  $(V_m, n_m, H_m)_{m\in \N}$ instead of $(V_m, n_m, H_m)_{m\in \No}$. We say that a \emph{financial position} $f$, i.e. a map $f:\mathcal{S}\rightarrow [-\infty,+\infty]$,
can be \emph{superhedged} by a (positive) generalized portfolio with initial endowment $V\in (-\infty,+\infty]$, if there is a (positive) generalized portfolio $(V_m, n_m, H_m)_{m\in \No}$
such that for every $S\in \mathcal{S}$
$$
f(S)\leq \sum_{m=0}^\infty \Pi^{V_m,n_m,H_m}_{\infty}(S) \;\textnormal{ and }\; V=\sum_{m=1}^\infty V_m.
$$
Note that the series above converge in $(-\infty,+\infty]$.

We now consider the superhedging operators
\begin{eqnarray*}
\bar I(f)&=&\inf\{V\in [0,+\infty]:\;f \textnormal{ can be superhedged by a positive   generalized} \\ &&\quad \textnormal{portfolio with initial endowment\;} V \}
\end{eqnarray*}
defined on the positive financial positions $f:\mathcal{S}\rightarrow [0,+\infty]$ and
\begin{eqnarray*}
\bar\sigma (f)&=&\inf\{V\in (-\infty,+\infty]:\;f \textnormal{ can be superhedged by a generalized} \\ &&\quad \textnormal{portfolio with initial endowment\;} V \}
\end{eqnarray*}
defined on all financial positions $f:\mathcal{S}\rightarrow [-\infty,+\infty]$. 

\begin{rem}
 The mapping $A\mapsto \bar I({\bf 1}_A)$ defined on the power set of $\mathcal{S}$ can be viewed as the analogue of Vovk's
 outer measure \cite{shafer, vovk1} in our context. We here already point to the subtlety that, depending on the choice 
 of the trajectory set $\mathcal{S}$, the superhedging operators $\bar \sigma$ 
 and $\bar I$ need not coincide on the set of positive financial positions, see Example \ref{exmp:I_neq_sigma} below.
 As $\bar I$ is countably sub-additive and $\bar \sigma$ is, in general, only finitely subadditive, 
 we shall define null sets in terms of $\bar I$, following the standard procedure in non-lattice integration \cite{leinert, konig} and in continuous time model-free finance, e.g. \cite{vovk1, perkowski, bartl}. However, by resorting to the theory of non-lattice integration in our financial context, we shall argue in the next section that, in general, $\bar \sigma$ 
 acts as the outer superhedging integral, which should be applied for option pricing, while 
 the role of $\bar I$ is to induce a norm which is used for the continuous extension of pricing beyond simple strategies.  
\end{rem}

We say that the trajectory set $\mathcal{S}$ admits a \emph{model-independent arbitrage} in the class of generalized portfolios, if there is a financial position $f$ such that $f(S)>0$ for every $S\in \mathcal{S}$
and such that $f$ can be superhedged by a generalized portfolio with initial endowment 0. If one can choose $f\equiv 1$, we will speak of a \emph{strict model-independent arbitrage}.
The rationale of a model-independent arbitrage is that it always persists, even if at a later stage
of the modelling process some trajectories in $\mathcal{S}$ are judged as `implausible' (e.g., by introducing null sets)
and arbitrage considerations are restricted to a reduced trajectory set.
We emphasize that several notions of `model-independent arbitrage' can be found in the literature. Our definition of model-independent arbitrage follows  the one in \cite{acciaio}, 
while our definition of strict model-independent arbitrage is closely related to the notion of model-independent arbitrage introduced in \cite{davis}. For a more detailed discussion on various notions of no arbitrage under model uncertainty in discrete time, we refer to  \cite{burzoni}.

The previous considerations suggest that absence of a strict model-inde\-pendent
arbitrage in the class of generalized portfolios is a minimal requirement for any `reasonable' trajectory set, 
if we accept the idealization of generalized portfolios. 
The following proposition provides a simple equivalent characterisation
in terms of the superhedging operator $\bar \sigma$.

\begin{prop}\label{prop:arbitrage}
 There is no strict model-independent arbitrage in the class of generalized portfolios, if and only if 
 \begin{eqnarray*}
(L)& \hspace{4cm}& \bar\sigma(0)\geq 0. \hspace{5cm} \
\end{eqnarray*}
\end{prop}
We refer to (L) as \emph{Leinert's condition}, as it turns out to be equivalent 
to the continuity condition imposed in his general theory of non-lattice integration \cite{leinert}, see Proposition \ref{prop:Leinert} below. Since the converse inequality $\bar\sigma(0)\leq 0$ is always satisfied, (L) is also equivalent to $\bar\sigma(0)=0$.
\begin{proof}
 Suppose first that $\bar\sigma(0)<0$. Then $0$ can be superhedged by a generalized trading strategy $(V_m, n_m, H_m)_{m\in \No}$ with initial endowment $-\epsilon$ for some $\epsilon>0$.
 After rescaling the portfolio
by the constant factor $1/\epsilon$, we may assume
that $\epsilon=1$. Hence, shifting $V_0$ by +1, we come up with a generalized portfolio which guarantees a payoff of at least 1 in every conceivable scenario with zero initial endowment.
Now suppose that there is a strict model dependent arbitrage $(V_m, n_m, H_m)_{m\in \No}$  in the class of generalized portfolios. Shifting $V_0$ by -1, we obtain a superhedge
for 0 in the same class with initial endowment -1. Hence, $\bar\sigma(0)\leq -1$.
\end{proof}

\begin{rem}\label{rem:constant}
 A simple sufficient condition for (L) is that the trajectory set $\mathcal{S}$ contains the constant trajectory $S^{\textnormal{const}}=(s_0,s_0,\ldots)$, because whenever 0 is
 superhedged by a generalized portfolio $(V_m, n_m, H_m)_{m\in \No}$, we obtain
 $$
 0\leq \sum_{m=0}^\infty \Pi^{V_m,n_m,H_m}_{\infty}(S^{\textnormal{const}})=\sum_{m=0}^\infty V_m,
 $$
 i.e. the initial endowment of such a superhedge is nonnegative.
\end{rem}

We close this section with some notation concerning conditional trajectory spaces, which is required to restart the model at a later time: Suppose $S\in \mathcal{S}$ and $j\in \mathbb{N}_0$. If the stock price follows the trajectory
$S$ up to time $j$, i.e. if the initial trajectory $(S_0,\ldots, S_{j})$ realizes, then the \emph{conditional  space}
$$
\mathcal{S}_{(S,j)}=\{\tilde S\in \mathcal{S};\; (\tilde S_0,\ldots \tilde S_j)=(S_0,\ldots,S_j) \}
$$
consists of all conceivable trajectories which are compatible with the realized initial segment of the trajectory. We occasionally refer to conditional spaces as \emph{nodes}, when the focus is on the  local (in time) behaviour of the trajectories in $\mathcal{S}_{(S,j)}$.

For a given conditional space $\mathcal{S}_{(S,j)}$, we may think of restarting the market at time $j$ in this
situation, i.e. we throw away the past by shifting the time indices
and obtain the \emph{shifted conditional space}
$$
\mathcal{S}^{(S,j)}=\{(\tilde S_{j+i})_{i\in \No};\; \tilde S \in \mathcal{S}_{(S,j)}\}.
$$
All constructions above and in the remainder of this paper can be done with a shifted conditional space  in place of the original trajectory set.
We say that the trajectory set $\mathcal{S}$ \emph{satisfies a property $Q$ at every node}, if this property is satisfied by 
$\mathcal{S}^{(S,j)}$ for every $S \in \mathcal{S}$ and $j\in \No$.
For instance, we say that $\mathcal{S}$ satisfies the Leinert condition at every node -- and write (nL) for this property --, if each of the shifted conditional spaces $\mathcal{S}^{(S,j)}$ satisfies (L).

\section{Non-lattice integration in the sense of Leinert}  \label{sec:leinert}

In the next two sections, we review some concepts of non-lattice integration in the context of hedging prices in model-free finance. In a nutshell, the starting point is a positive linear functional
(`elementary integral')
on a vector space of `elementary integrands', which is then to be extended to  a larger `$L^1$-space'. The extension step can be done by exploiting continuity of the elementary integral
with respect to some appropriate semi-norm as in Leinert's approach or by introducing the notions of an outer and an inner integral as in K\"onig's approach. The crucial difficulty
compared to the classical Daniell theory is that the space of elementary integrands may fail to have the lattice property.

In our financial context, we consider the space of all financial positions which can be perfectly hedged by a simple portfolio
$$
\mathcal{E}=\{f=\Pi^{V,n,H}_\infty:\quad (V,n,H) \textnormal{ simple portfolio}\}
$$
as the space of elementary integrands and denote the subset of the positive positions in $\mathcal{E}$ by
$$
\mathcal{E}^+=\{f\in \mathcal{E}:\quad f(S)\geq 0 \textnormal{ for every } S\in \mathcal{S}\}.
$$
Clearly, $\mathcal{E}$ forms a vector space, but it fails to have the lattice property in typical situations except in binary models, as illustrated by
the following example.
\begin{exmp}\label{examp:no_lattice}
 Suppose that the range of stock prices at time $1$ contains at least three different values, i.e. there are trajectories $S^{(-)}, S^{(0)}, S^{(+)}\in \mathcal{S}$ such that
 $S^{(-)}_1< S^{(0)}_1< S^{(+)}_1$. Consider $f\in \mathcal{E}$ given by
 $$
 f(S)=(S_1- S^{(0)}_1)= (S_0-S^{(0)}_1) + (S_1-S_0),\quad S\in \mathcal{S}.
 $$
 We now assume that $|f|\in \mathcal{E}$, i.e. there is a simple portfolio $(V,n,G)$
 such that
 $$
 |f(S)|=V+\sum_{i=0}^{n-1} G_i(S)(S_{i+1}-S_i),\quad S\in \mathcal{S}
 $$
 We distinguish two cases: \\[0.1cm]
 (1) We may take $n=1$, i.e. for every $S\in \mathcal{S}$:
 $
 |f(S)|=V+G_0(S_{1}-s_0).
 $
   Then,
 $$
 0=|f(S^{(0)})|=V+G_0(S^{(0)}_1-s_0)
 $$
 and, consequently,
 $$
 |f(S)|=G_0(S_1-S^{(0)}_1),\quad S\in \mathcal{S}.
 $$
 If $G_0\neq 0$, then $G_0(S_1-S^{(0)}_1)<0$ for one of the choices $S\in\{S^{(-)},S^{(+)}\}$, a contradiction. If $G_0=0$, then $f(S^{(-)}_1)=0$ and, thus, $S^{(-)}_1= S^{(0)}_1$,
 which also leads to a contradiction.
 \\[0.1cm]
 (2) There is an $S^*\in \mathcal{S}$ such that
 $
 |f(S^*)|\neq V+G_0(S^*_{1}-S^*_{0}).
 $
 Let $a:=|f(S^*)|-(V+G_0(S^*_{1}-S^*_{0})).$ In the shifted conditional space $\mathcal{S}^{(S^*,1)}$, consider the simple portfolio $(0,n-1,H)$, where
 $$
 H_i(\tilde S_0,\ldots, \tilde S_i)= \frac{1}{a}G_{i+1}(S^*_0,\tilde S_0,\ldots \tilde S_{i}),\quad \tilde S\in \mathcal{S}^{(S^*,1)}, \;i=0,\ldots n-1.
 $$
 Since $(S^*_0,\tilde S_0,\ldots \tilde S_{n-1})=(S_0,\ldots,S_n)$ for some $S\in \mathcal{S}$ and $\tilde S_0=S^*_1$, the terminal wealth of this simple portfolio equals
 $$
 \frac{1}{a} \sum_{i=1}^{n-1}  G_{i}(S^*_0,\tilde S_0,\ldots \tilde S_{i-1})(\tilde S_{i}-\tilde S_{i-1})=\frac{1}{a}(|f(S^*)|- V+G_0(S^*_{1}-S^*_{0}))=1
 $$
 for every $\tilde S\in \mathcal{S}^{(S^*,1)}$. Hence, the shifted conditional space $\mathcal{S}^{(S^*,1)}$ is not free of strict model-independent arbitrage with simple strategies.
 \\[0.1cm]
 In conclusion, $\mathcal{E}$ can only have the lattice property, if $\mathcal{S}$ fails to 
 have the property of being free of strict model-independent arbitrage at every node with simple strategies,  or if the trajectory
 set has a binary structure.
\end{exmp}

Suppose now that the following trajectory-wise version of the \emph{law of one price}  holds for simple portfolios: 

\begin{itemize}
 \item[(LOP)\quad] If $(V_m, n_m, H_m)$, $m\in\{0,1\}$, are simple portfolios such that
 $$
 \Pi^{V_1,n_1,H_1}_{\infty}(S)=\Pi^{V_0,n_0,H_0}_{\infty}(S)
 $$
 for every $S\in \mathcal{S}$, then $V_1= V_0$.
\end{itemize}

Then, the hedging price functional  for elements in $\mathcal{E}$ (with trading  restricted to simple portfolios) is a well-defined linear operator:
$$
I:\mathcal{E}\rightarrow \R, \quad \Pi^{V,n,H}_\infty \mapsto V.
$$
If, moreover, $I$ is a positive functional, i.e. $I(f)\geq 0$ for every $f\in \mathcal{E}^+$, we can re-write
\begin{eqnarray}
\label{eq:barI} \bar I(f)&=&\inf\{\sum_{m=1}^\infty I(f_m):\; (f_m)_{m\in \N} \textnormal{ in }\mathcal{E}^+,\; \sum_{m=1}^\infty f_m\geq f\}, \\
 \label{eq:barsigma} \bar \sigma(f) &=& \inf\{\sum_{m=0}^\infty I(f_m):\; f_0\in \mathcal{E},\;(f_m)_{m\in \N} \textnormal{ in }\mathcal{E}^+,\; \sum_{m=1}^\infty f_m\geq f\}.
\end{eqnarray}

The following elementary lemma will be used in the characterisation of Leinert's condition (L).

\begin{lem}\label{lem:mon}
 (LOP) holds and $I$ is positive, if and only if the following monotonicity condition is in force:
 \begin{itemize}
 \item[(MON)\quad]  If $(V_m, n_m, H_m)$, $m\in\{0,1\}$, are simple portfolios such that
 $$
 \Pi^{V_1,n_1,H_1}_{\infty}(S)\geq \Pi^{V_0,n_0,H_0}_{\infty}(S)
 $$
 for every $S\in \mathcal{S}$, then $V_1\geq V_0$
\end{itemize} 
\end{lem}

We refer to \cite{ferrando, bender} for an easy-to-check sufficient condition for (MON) in terms of the local behaviour of the trajectories, which is called local-zero neutrality.

\begin{prop}\label{prop:Leinert}
 The following assertions are equivalent:
 \begin{enumerate}
  \item The Leinert condition (L) holds.
  \item The trajectory-wise (LOP) holds,  the hedging price functional $I$ is positive and satisfies the following continuity condition: For every $f\in \mathcal{E}$
  $$
  I(f)\leq \bar I(f^+)
  $$
   \end{enumerate}
Moreover, $\bar\sigma(f)=I(f)$ for every $f\in \mathcal{E}$ and $\bar I(f)=I(f)$ for every $f\in \mathcal{E}^+$, if one (and then both) of these two equivalent conditions holds.
\end{prop}
\begin{proof}
 `1. $\Rightarrow$ 2.': We suppose that (L) is in force, i.e.  $\bar\sigma(0)\geq 0$. 
 We first verify the monotonicity condition (MON): To this end, we may assume without loss of generality that $n_0=n_1$ (by passing to $\max\{n_0,n_1\}$ if necessary). Consider the simple strategy $(V,n,H)=(V_1-V_0,n_0,H_1-H_0)$  which satisfies
 $$
 \Pi^{V,n,H}_{\infty}(S)=\Pi^{V_1,n_1,H_1}_{\infty}(S)- \Pi^{V_0,n_0,H_0}_{\infty}(S)\geq 0
 $$
 for every $S\in \mathcal{S}$. Hence, $V=V_1-V_0\geq \bar\sigma(\Pi^{V,n,H}_{\infty})\geq \bar\sigma(0)\geq 0$.
 Consequently, by Lemma \ref{lem:mon}, the law of one price holds and $I$ is positive. 
 We now turn to the representations of $I$ in terms of $\bar \sigma$ and $\bar I$ and to the continuity condition:
 Let $f\in \mathcal{E}$ and consider any $f_0 \in \mathcal{E}$, $(f_m)_{m\in \N}$ in $\mathcal{E}^+$
 such that $f\leq \sum_{m=0}^\infty f_m$ on $\mathcal{S}$, or equivalently, with $f'_0:=f_0-f$,
 $$
 0 \leq f'_0(S) +\sum_{m=1}^\infty f_m(S)
 $$
 for every $S\in \mathcal{S}$. Hence,
 $$
 I(f'_0)+\sum_{m=1}^\infty I(f_m) \geq \bar\sigma(0)\geq 0.
 $$
 By linearity of $I$,
 $$
 I(f)\leq \sum_{m=0}^\infty I(f_m)
 $$
 Taking the infimum over the $f_m$'s (with the properties specified above) yields
 $$
 I(f)\leq \bar\sigma(f).
 $$
 For the other inequality, choose $f_0=f$ and $f_m=0$ for $m\in \N$, implying
 $$
 \bar\sigma(f)\leq \sum_{m=0}^\infty I(f_m)= I(f).
 $$
 Hence, the property $I(f)=\bar\sigma(f)$ for every $f\in \mathcal{E}$ is established under the assumption (L). Then,
 the continuity condition in 2.
 follows by
 $$
 I(f)=\bar \sigma(f) \leq \bar\sigma (f^+)\leq \bar I(f^+).
 $$
 Moreover, for every $f\in \mathcal{E}^+$,
 $$
 I(f)=\bar \sigma(f)\leq \bar I(f)\leq I(f).
 $$
\\[0.2cm]
`2. $\Rightarrow$ 1.':
 Suppose $f_0\in \mathcal{E}$ and $(f_m)_{m\in \N}$ in $\mathcal{E}^+$ such that
 $
 0\leq \sum_{m=0}^\infty f_m.
 $
 Then, applying the continuity condition for $f=-f_0$,
 $$
 I(f)\leq \bar I(f^+)\leq \bar I( \sum_{m=1}^\infty f_m)\leq \sum_{m=1}^\infty I(f_m).
 $$
 Linearity of $I$ implies
 $$
 \sum_{m=0}^\infty I(f_m)\geq 0.
 $$
 Passing to the infimum, we obtain $\bar\sigma(0)\geq 0$.
\end{proof}

We now extend the hedging functional $I$ to a wider class of financial positions following \emph{Leinert's construction of a non-lattice integral} \cite{leinert} under the
assumption (L):
\begin{enumerate}
 \item Define for any financial position $f:\mathcal{S}\rightarrow [-\infty,+\infty]$
 \begin{equation}\label{eq:integralnorm}
 \|f\|:=\bar I(|f|)
 \end{equation}
 and call $f$ a \emph{null function}, if $\|f\|=0$. A \emph{null set} is any set $A\subset \mathcal{S}$ such that the indicator function of $A$ is a null function, and a property is said to hold
 almost everywhere (a.e.), if it is valid outside a null set. Thanks to the countable subadditivity of $\bar I$, we can manipulate null functions and null sets in the
 usual way.
 \item The space $\mathcal{F}$ of all real-valued financial positions $f:\mathcal{S}\rightarrow (-\infty,+\infty)$ such that $\|f\|<\infty$ can be shown to be complete endowed with
 the semi-norm $\|\cdot\|$. (It, thus, becomes a Banach space, if we identify functions which only differ by a null function).
 \item Denote by $\mathcal{E}'$ the space of financial positions $f$ in $\mathcal{E}$ such that $\|f\| <\infty$. As (L) holds, Proposition \ref{prop:Leinert} implies that
 the restriction of the hedging price functional $I$ to  $\mathcal{E}'$ is continuous with respect to $\|\cdot\|$ and thus uniquely extends to a continuous linear functional defined
 on the closure of $\mathcal{E}'$ with respect to the norm $\|\cdot\|$.
 \item Denoting this closure by $\mathcal{L}^1_{(L)}$, we call this unique extension (in our particular financial context) the  \emph{hedging price integral in the sense of Leinert}:
 $$
 \int_{(L)}: \mathcal{L}^1_L\rightarrow \R,\quad f \mapsto  \int_{(L)} f
 $$
 It satisfies, for every $f\in \mathcal{L}^1_{(L)}$,
 \begin{equation}
 \int_{(L)} f = \bar\sigma(f) \label{eq:outer_integral},
 \end{equation}
 i.e. the hedging price integral in the sense of Leinert is the restriction of the superhedging operator $\bar\sigma$ to the linear subspace  $\mathcal{L}^1_{(L)}$.
 Note that, by construction, any null function $h$ belongs to $\mathcal{L}^1_{(L)}$ and $\int_{(L)} h=0$. We  write
 $(\mathcal{L}^1_{(L)})^+$ for the subset of positive financial positions in $\mathcal{L}^1_{(L)}$.
\end{enumerate}

Equations (\ref{eq:integralnorm})--(\ref{eq:outer_integral}) explain, why we should think of the superhedging operator $\bar I$ as a norm and of $\bar \sigma$
as an outer integral, respectively.

We collect some  properties of the hedging price integral, which follow immediately by the construction and by Proposition \ref{prop:Leinert}.
\begin{prop}
 Suppose (L). Then the hedging price integral $\int_{(L)}: \mathcal{L}^1_L\rightarrow \R$ is a linear, positive, continuous (w.r.t. $\|\cdot\|$), and constant-preserving operator.
\end{prop}

The hedging price integral, thus, always shares all the properties of a \emph{linear price system} induced by an equivalent martingale measure in the classical probabilistic framework except for the strict positivity which is formulated with respect to
the null sets of the reference probability measure in the probabilistic framework. A model-free version of strict positivity can be formulated with reference
to the null functions of the semi-norm $\|\cdot\|$ in the following way.
\begin{defi}
 Suppose (L). The hedging price integral $\int_{(L)}$ is said to be \emph{strictly positive}, if for every $f\in (\mathcal{L}^1_{(L)})^+$
 $$
  \int_{(L)} f=0 \quad \Rightarrow \quad f \textnormal{ is a null function}.
 $$
\end{defi}

\begin{rem}\label{rem:strict_positivity}
 Suppose that $\int_{(L)}$ is strictly positive and that $f,g\in \mathcal{L}^1_{(L)}$ are two financial positions such that $f\geq g$ almost everywhere (w.r.t. to the null sets of $\|\cdot\|$),
 but which
are assigned the same price by the hedging price integral, i.e. $\int_{(L)} f = \int_{(L)} g$. We are going to show that $f=g$ a.e., which is the usual way to formulate strict
positivity of a price system.

As the positive part $(g-f)_+$ of $g-f$ is then a null function, we observe
that $h=f-g+(g-f)_+ \in  (\mathcal{L}^1_{(L)})^+$ and
$$
\int_{(L)} h = \int_{(L)} f - \int_{(L)} g=0
$$
By strict positivity, $h$ is a null function and then so is $f-g$, because $\|f-g\|\leq \|h\|+ \|(g-f)_+\|=0$. Thus, $f=g$ a.e.
\end{rem}

With a notion of negligibility in terms of $\|\cdot\|$-null sets at hand, we can also define the following notion  of arbitrage, cp. \cite{cassese}.
\begin{defi}
A generalized portfolio $(V_m,n_m,H_m)_{m\in \N_0}$ is said to be an \emph{arbitrage with respect to the $\|\cdot\|$-null sets}, if 
 $
 \sum_{m=0}^\infty V_m=0,
 $
  $$
 \left\{S:\; \sum_{m=0}^\infty \Pi^{V_m,n_m,H_m}_\infty(S) <0  \right\} 
 $$
 is a null set (w.r.t. $\|\cdot\|$) and 
 $$
 \left\{S:\; \sum_{m=0}^\infty \Pi^{V_m,n_m,H_m}_\infty(S) >0  \right\} 
 $$
 is not a null set (w.r.t. $\|\cdot\|$). 
 \end{defi}
In the special case of a simple portfolio,  this notion of an arbitrage with respect to the $\|\cdot\|$-null sets simplifies as follows:
A simple portfolio $(V,n,H)$ is an arbitrage with respect to the $\|\cdot\|$-null sets, if $V=0$,
 $\Pi^{V,n,H}_\infty \geq 0$ almost everywhere, and if $\{S:\; \Pi^{V,n,H}_\infty(S)>0 \}$ 
 is not a null set.
This notion of arbitrage is analogous to the classical one, replacing the null sets of the reference probability measure by the null sets induced by the superhedging functional $\bar I$.

The following example demonstrates that the Leinert condition (L) alone is neither sufficient to guarantee the strict positivity of the hedging price integral nor to obtain absence of arbitrage with respect to the $\|\cdot\|$-null sets.

\begin{exmp}\label{exmp:I_neq_sigma}
 We consider the following trajectory set: Trajectories start at value 1. Between time 0 and time 1 they can move either up or down. If they move up, they have value
 2 at time 1. At later times, trajectories in this up-node behave as follows: They stay constant at value 2 for some time but eventually jump up to 4. If trajectories
 move down between time 0 and 1, they can attain any value $1-1/n$, $n\geq 1$, at time 1 and stay at that value forever. Formally
 $$
 \mathcal{S}=\{S^{(u,n)}, S^{(d,n)}:\quad n\in \mathbb{N} \}
 $$
 where
 $$
 S^{(u,n)}_0=1,\quad S^{(u,n)}_1=2,\quad S^{(u,n)}_i=2 \textnormal{ for $i<n+1$, and } S^{(u,n)}_i=4 \textnormal{ for $i\geq n+1$}
 $$
 and
 $$
 S^{(d,n)}_0=1,\quad S^{(d,n)}_i=1-1/n \textnormal{ for $i\geq 1$}.
 $$
 Notice that $\mathcal{E}$ is not a vector lattice for this trajectory set $\mathcal{S}$ by Example \ref{examp:no_lattice}.
 We denote the `up-node' at time 1 by $\mathcal{S}_{u}=\{S^{(u,n)}, \quad n\in \mathbb{N} \}=\{S\in \mathcal{S}:\; S_1>S_0 \}$ and the `down-branch' at time 1
 by $\mathcal{S}_{d}=\{S^{(d,n)}, \quad n\in \mathbb{N} \}=\{S\in \mathcal{S}:\; S_1<S_0 \}$.
 \\[0.1cm]
 (1) We first show that the Leinert condition $\bar \sigma(0)\geq 0$ is satisfied:
   Consider $\Pi^{V_m,n_m,H_m}_\infty \in \mathcal{E}^+$, $m\in \N$, and $\Pi^{V_0,n_0,H_0}_\infty \in \mathcal{E}$ such that
  $$
  F:=\sum_{m=0}^\infty  \Pi^{V_m,n_m,H_m}_\infty \geq 0.
  $$
 We need to show that $\sum_{m=0}^\infty V_m\geq 0$, for which we may and do assume that $\sum_{m=1}^\infty V_m<+\infty$.
 On the one hand, note that for every $m\geq 1$
 $$
 V_m+H_{m,0}=V_m+ H_{m,0}(S^{(u,n_m)}_1-S^{(u,n_m)}_0)= \Pi^{V_m,n_m,H_m}_\infty(S^{(u,n_m)})\geq 0,
 $$
 as the trajectory $S^{(u,n_m)}$ stays constant between times 1 and $n_m$. Hence, $\sum_{m=0}^\infty H_{m,0}$ converges in $(-\infty,+\infty]$.
 On the other hand,  all trajectories in the down-branch stay constant after
 time 1. Hence
 \begin{equation}\label{eq:0005}
 0\leq F(S^{d,n})=\sum_{m=0}^\infty\left( V_m-  \frac{H_{m,0}}{n} \right),\quad n\in \mathbb{N},
 \end{equation}
 (including convergence of the series for every $n\in \N$, possibly with value $+\infty$). Thus, $\sum_{m=0}^\infty H_{m,0}$ converges in $[-\infty,+\infty)$. In conclusion,
 $\sum_{m=0}^\infty H_{m,0}$ converges in the reals, and \eqref{eq:0005} can then be rewritten as
$$
\sum_{m=0}^\infty V_m \geq  \frac{1}{n} \sum_{m=0}^\infty H_{m,0}.
$$
 Passing with $n$ to infinity, implies that $\sum_{m=0}^\infty V_m\geq 0$ and concludes the proof of the Leinert condition (L).
 \\[0.1cm]
 (2)
We now consider a straddle with maturity $T=1$ struck at $K=1$, i.e.
$$
g(S)=|S_1-1|,\quad S\in \mathcal{S},
$$
 which is strictly positive on $\mathcal{S}$. We claim that $g\in \mathcal{L}^1_{(L)}$ and  $\int_{(L)} g=\bar\sigma(g)=0$, i.e. the hedging price integral
 trades this (strictly positive) straddle for a price of zero, which does not make sense from a financial point of view.

 In order to show these assertions, we consider
 $$
 g_1(S)=-(S_1-S_0),\quad g_2(S)= 2\cdot {\bf 1}_{\mathcal{S}_u}(S),\quad S\in \mathcal{S},
 $$
 which satisfy $g=g_1+g_2$. Obviously, $g_1\in \mathcal{E}'$ with $\bar\sigma(g_1)=I(g_1)=0$. Hence, it is sufficient to show that $g_2$ is a null function. We can write, however,
 $$
 g_2= \sum_{m=1}^\infty (S_{m+1}-S_m)=: \sum_{m=1}^\infty f_m
 $$
 where each $f_m$ belongs to $\mathcal{E}^+$ (as it takes values in $\{0,2\}$ only) with $I(f_m)=0$. Thus, $\|g_2\|=\bar I(g_2)=0$.
 \\[0.1cm]
 (3) We will next show that $\bar I(g)>0$ and, hence, $g$ is not a null function. This completes the proof that the hedging price integral is not strictly
 positive in this example. Suppose on the contrary, that  $\bar I(g)=0$. Then, by Proposition \ref{prop:Leinert} and by the countable subadditivity of $\bar I$
 $$
 1=I(1) = \bar I(1)\leq \bar I(\sum_{m=1}^\infty g)\leq \sum_{m=1}^\infty  \bar I(g)=0,
 $$
 a contradiction.
 \\[0.2cm] 
 (4) Let us reconsider the findings of this example from an arbitrage point of view.  We claim that, in this example, the simple portfolio with zero initial endowment, which shortens one share of stock up to time 1 ($V=0$, $n=1$, $H_0=-1$) is an arbitrage with respect to the $\|\cdot\|$-null sets. By part (2) of this example,
 $\mathcal{S}_u$ is a null set, while for $S\in \mathcal{S}_d$ the terminal wealth of this 
 simple portfolio is 
 $
 S_0-S_1>0.
 $
 Finally, the set $\mathcal{S}_d$ is not a null set, since
 $$
 1=I(1)=\bar\sigma(1) \leq \bar \sigma({\bf 1}_{\mathcal{S}_d}) + \bar \sigma({\bf 1}_{\mathcal{S}_u}) =\bar \sigma({\bf 1}_{\mathcal{S}_d}) \leq \bar I({\bf 1}_{\mathcal{S}_d}) \leq 1.
 $$
 Here, we applied Proposition \ref{prop:Leinert} for the second identity and note that
 $ \bar \sigma({\bf 1}_{\mathcal{S}_u})=0$, because
 $0\leq \bar\sigma(0) \leq \bar \sigma({\bf 1}_{\mathcal{S}_u})\leq \bar I({\bf 1}_{\mathcal{S}_u})=0$.
  \end{exmp}

From a mathematical point of view, the previous example illustrates the inconsistency that the superhedging operators $\bar I$ and $\bar\sigma$ may fail to coincide
on the set $(\mathcal{L}^1_{(L)})^+$. This is obviously the only way, how to break the strict positivity of the hedging price integral:
\begin{prop}\label{prop:strict_positivity}
 Suppose (L). If $\bar I(f)=\bar \sigma(f)$ for every $f\in (\mathcal{L}^1_{(L)})^+$, then the hedging price integral $\int_{(L)}: \mathcal{L}^1_L\rightarrow \R$ is strictly positive
 (and, thus, qualifies as a linear price system).
\end{prop}
\begin{proof}
 The proof is immediate, since then
 $$
 \int_{(L)} f =\bar\sigma(f)=\bar I(f)=\|f\|
 $$
 for every $f\in (\mathcal{L}^1_{(L)})^+$.
\end{proof}

\section{Non-lattice integration in the sense of K\"onig}\label{sec:konig}

The discussion in the previous section  motivates  to  strengthen Leinert's continuity condition in a way that the superhedging operators
$\bar I$ and $\bar \sigma$ coincide on a sufficiently rich class of positive financial positions, in order to obtain a sound interpretation of the
integral as a linear price system. This problem has been addressed by K\"onig \cite{konig} in the general theory of non-lattice integration.

In our context, we say that a financial position $f:\mathcal{S}\rightarrow [-\infty,+\infty]$ has \emph{maturity} $n$, if  $f$
only depends on $S$ through
the initial segment $(S_0\ldots, S_n)$. $f$ is said to have \emph{finite maturity}, if it has maturity $n$ for some $n\in \mathbb{N}_0$.

\begin{prop}\label{prop:Koenig}
 The following assertions are equivalent:
 \begin{enumerate}
  \item $\bar I(f)=\bar \sigma(f)$ for every positive financial position $f:\mathcal{S}\rightarrow [0,+\infty]$.
  \item $\bar I(f)=\bar \sigma(f)$ for every positive and real valued financial position $f:\mathcal{S}\rightarrow [0,+\infty)$ with finite maturity.
  \item (LOP) and K\"onig's continuity condition
  \begin{eqnarray*}
  (K)& \hspace{2cm}& I(f) +\bar I(f^-) \leq \bar I(f^+),\quad  f\in \mathcal{E} \hspace{5cm} \
  \end{eqnarray*}
  hold.
 \end{enumerate}
 Moreover, each of the three conditions implies (L).
\end{prop}

\begin{proof}
 `3. $\Rightarrow$ 1.': We first show that (K) implies that $I$ is a positive functional. Indeed, for $f\in \mathcal{E}$, we may apply (K) to $f$ and $-f$
 in order to obtain
 $$
 I(f) +\bar I(f^-) = \bar I(f^+).
 $$
 If $f\in \mathcal{E}^+$, the previous equality yields $I(f)=\bar I(f)\geq 0$. Given the positivity of $I$, the implication `3. $\Rightarrow$ 1.' has been shown by K\"onig \cite{konig} in a more general context.
 \\[0.1cm]
 The implication `1. $\Rightarrow$ 2.' is trivial. 
 \\[0.1cm]
 It remains
 to prove that 2. implies (L) and 3.: 
 Applying 2. to the function $f \equiv 0$ yields $\bar\sigma(0)=\bar I(0)=0$, and, thus, (L). However, (L) implies (LOP) by Proposition \ref{prop:Leinert}.
 In order to prove (K), we may and do assume that $\bar I(f^+)<\infty$ and denote by
 $(V_m, n_m, H_m)_{m\in \N}$ a positive generalized portfolio which superhedges $f^+$. As $-f=\Pi^{V_0,n_0,H_0}_\infty$ for some simple portfolio $(V_0,n_0,H_0)$, the generalized
 portfolio $(V_m, n_m, H_m)_{m\in \No}$ is a superhedge for $f^-$ and, thus,
 $$
 \bar\sigma(f^-) \leq -I(f)+\bar I(f^+) <\infty.
 $$
 Noting that $f^-$ inherits the finite maturity from $f$, we obtain $ \bar\sigma(f^-)= \bar I(f^-)$, and the proof of (K) is complete.
\end{proof}

\begin{rem}
 The previous Proposition shows that (LOP) and (K) imply (L). However, (L) does not imply (K) in general. Example \ref{exmp:I_neq_sigma} 
 applies as a counterexample.
\end{rem}

We next review \emph{K\"onig's construction of a non-lattice integral}  \cite{konig} in our context and, for the moment, only assume that (L) is in force:
\begin{enumerate}
 \item Recall that $\bar \sigma$ can be interpreted as an outer integral, which motivates to consider the inner integral
 $$
 \underline \sigma(f)=-\bar\sigma(-f)
 $$
 for every financial position $f:\mathcal{S}\rightarrow [-\infty,+\infty]$. Then, (L) ensures that the inequality
 $\underline \sigma(f)\leq \bar \sigma(f)$ is valid for every financial position $f$ by \cite[Bemerkung 1.5]{konig}, where both sides of the inequality may be $+\infty$.
 \item We write $\mathcal{L}^1_{(K)}$ for the space of all real-valued financial positions $f$ such that $\underline \sigma(f)= \bar \sigma(f)\in (-\infty,+\infty)$. Then, the restriction
 of $\bar\sigma$ to $\mathcal{L}^1_{(K)}$ is a linear mapping \cite[Behauptung 2.1]{konig}, which we denote by
 $$
 \int_{(K)}: \mathcal{L}^1_{(K)}\rightarrow \mathbb{R},\quad f\mapsto \bar\sigma(f).
 $$
 In our context, we call it the \emph{hedging price integral in the sense of K\"onig}.
 \item Note that $\mathcal{E}\subset \mathcal{L}^1_{(K)}$ by Proposition \ref{prop:Leinert} and that $\mathcal{L}^1_{(L)} \subset \mathcal{L}^1_{(K)}$ by the vector space property of $\mathcal{L}^1_{(L)}$. Hence, K\"onig's integral construction
 extends Leinert's one to a possibly larger class of integrands.
\end{enumerate}

If the continuity condition (K) is imposed, K\"onig's integral satisfies the following properties.

\begin{thm}\label{thm:Koenig}
 Assume (LOP) and (K). Then, $\int_{(K)}: \mathcal{L}^1_{(K)}\rightarrow \mathbb{R}$ is a linear, positive, constant-preserving extension of $I$, which satisfies the following continuity condition:  If
 $f,f_n \in \mathcal{L}^1_{(K)}$, then
 $$
 \lim_{n\rightarrow \infty} \|f_n-f\|=0\quad \Rightarrow \quad \lim_{n\rightarrow \infty} \int_{(K)} f_n= \int_{(K)} f.
 $$
 Moreover, $\int_{(K)}$ is strictly positive in the following sense: If $f,g\in \mathcal{L}^1_{(K)}$ are two financial positions such that
 $f\geq g$ almost everywhere (w.r.t. to the null sets of $\|\cdot\|$),
 but which
are assigned the same price by the hedging price integral, i.e. $\int_{(K)} f = \int_{(K)} g$, then $f=g$ almost everywhere.

Hence, the hedging price integral in the sense of K\"onig qualifies as a linear price system.
\end{thm}
\begin{proof}
 The first three properties are elementary. Continuity is stated in \cite{konig}, Satz 2.9. Finally, the strict positivity is a consequence
 of Proposition \ref{prop:Koenig} following the arguments in Proposition \ref{prop:strict_positivity} and Remark \ref{rem:strict_positivity}.
\end{proof}
The domain of the non-lattice integral $\int_{(K)}$ under assumption (K) has been characterized in \cite{konig}.
Denote by $\mathfrak{M}$ the class of financial positions $f$, which can be written in the form $f=\sum_{m=1}^\infty f_m$ where $(f_m)_{m\in\N}$ is
 a sequence in $\mathcal{E}^+$ and $\sum_{m=1}^\infty I(f_m)<\infty.$
\begin{thm}\label{thm:K_L_1}
 Assume (LOP) and (K). Then, for every real-valued financial position $f$ the following assertions are equivalent:
 \begin{enumerate}
  \item $f\in \mathcal{L}^1_{(K)}$.
  \item There is a sequence $(f_m)$  in $\mathcal{E}$ such that
  $$
  \lim_{m\rightarrow \infty} \|f_m-f\|=0.
  $$
  \item $f$ has a representation of the form
  $$
  f=g+h_1-h_2\quad a.e.,
  $$
  where $g\in \mathcal{E}$ and $h_1,h_2\in \mathfrak{M}$.
 \end{enumerate}
\end{thm}

We close this section by observing that K\"onig's condition rules out arbitrage 
with respect to the $\|\cdot\|$-null sets.
\begin{thm}\label{thm:arbitrage_K}
 Suppose (LOP) and (K) are in force. Then there is no arbitrage with respect to the $\|\cdot\|$-null sets in the class of generalized portfolios.
\end{thm}
\begin{proof}
 Suppose that $(V_m, n_m, H_m)_{m\in \No}$ is a generalized portfolio with zero initial endowment, i.e.
 $V:=\sum_{m=0}^\infty V_m=0$. By the previous theorem, the terminal wealth of 
 this generalized portfolio $V_\infty:=\sum_{m=0}^\infty \Pi^{V_m,n_m,H_m}_\infty$ belongs to $\mathcal{L}^1_{(K)}$ and $\int_{(K)} V_\infty=0$ by the continuity property in Theorem \ref{thm:Koenig}, since for $M\geq 1$
 $$
 \| V_\infty-\sum_{m=0}^M \Pi^{V_m,n_m,H_m}_\infty \|\leq \sum_{m=M+1}^\infty \bar I(\Pi^{V_m,n_m,H_m}_\infty)=\sum_{m=M+1}^\infty V_m \rightarrow 0\; (M\rightarrow \infty).
 $$ If $V_\infty\geq 0$ almost everywhere (w.r.t. to the null sets of $\|\cdot\|$), then $V_\infty$ is a null function by the strict positivity in Theorem \ref{thm:Koenig}, and, thus, $\{S:\; V_\infty(S)>0\}$ is a null set. Hence, $(V_m, n_m, H_m)_{m\in \No}$ is not an arbitrage with respect to to the $\|\cdot \|$-null sets.
\end{proof}

\section{On the relation between Leinert's and K\"onig's continuity condition}\label{sec:L_vs_K}

Theorem \ref{thm:Koenig} shows that K\"onig's continuity condition (K) is crucial to guarantee a meaningful interpretation of the hedging price integral as a linear
price system on $\mathcal{L}^1_{(K)}$. Moreover, by Theorem \ref{thm:arbitrage_K}, condition (K) ensures that a generalized portfolio can never be an arbitrage with respect to the null sets induced by the superhedging functional $\bar I$.
However, Example \ref{exmp:I_neq_sigma} illustrates that K\"onig's continuity condition is not implied by Leinert's condition in general in our hedging and pricing context.
Thus, the remaining sections are devoted to the derivation of sufficient conditions for (K). As a first key result in this direction, we
prove in this section  that the Leinert condition at every node implies the K\"onig condition at every node, and, thus, in particular (K) for the original
trajectory set $\mathcal{S}$. We shall write (nL) and (nK) for the nodewise versions of the two continuity conditions, i.e. when property (L), resp. (K) is satisfied by the shifted conditional spaces
$\mathcal{S}^{(S,j)}$ for every $S \in \mathcal{S}$ and $j\in \No$.

\begin{thm}\label{thm:cond_Koenig}
 The Leinert condition  at every node (nL)  implies the K\"onig condition  at every node (nK).
\end{thm}

In combination with Remark \ref{rem:constant}, we arrive at the first simple sufficient condition for (K). More general criteria will be provided in the next section.
\begin{cor}
 Conditions (LOP) and (K) hold, if, for every $S\in \mathcal{S}$ and $j\in \mathbb{N}_0$, the stopped trajectory $(S_{i\wedge j})_{i \in \N_0}$ belongs to $\mathcal{S}$.
\end{cor}
\begin{rem}
  Bartl et al. \cite{bartl} argue that many relevant trajectory spaces 
  in continuous-time model-free finance are closed against stopping of trajectories. In their continuous-time setting, they impose the condition that stopped trajectories again belong 
  to the trajectory space in order to avoid additional admissibility conditions on the portfolios, cp. also Remark 2.15 in \cite{bartl2}.
\end{rem}

The proof of Theorem \ref{thm:cond_Koenig} requires several preparations. We start with some definitions concerning the classifications of nodes, cp. \cite{ferrando}.
\begin{defi}
 Given a  fixed trajectory set $\mathcal{S}$, we say that a node $\mathcal{S}_{(S,j)}$ is an \emph{arbitrage node}, if there is an $\varepsilon \in \{-1,1\}$ such that
 for every $\tilde S\in \mathcal{S}_{(S,j)}$
 $$
 \varepsilon (\tilde S_{j+1}-S_j)\geq 0,
 $$
 and the strict inequality holds for at least one $\tilde S'\in \mathcal{S}_{(S,j)}$. An arbitrage node is called \emph{type I}, if there
 is an $S'\in \mathcal{S}_{(S,j)}$ such that $S'_{j+1}=S_j$. Otherwise it is called \emph{type II}. A node $\mathcal{S}_{(S,j)}$ is said to be \emph{flat}, if
 $\tilde S_{j+1}=S_j$ for every $\tilde S\in \mathcal{S}_{(S,j)}$. Nodes which are neither flat nor arbitrage nodes will be called \emph{up-down nodes}.
\end{defi}

We denote
 \begin{eqnarray*}
 \mathcal{N}&:=&\{S\in \mathcal{S}:\quad \mathcal{S}_{(S,j)} \textnormal{ is an arbitrage node of type I} \\ && \quad \textnormal{ and }S_{j+1}\neq S_j
  \textnormal{ for some }j\in \No\}. \\
  \mathcal{N}_n&:=&\{S\in \mathcal{S}:\quad \mathcal{S}_{(S,j)} \textnormal{ is an arbitrage node of type I} \\ && \quad \textnormal{ and }S_{j+1}\neq S_j
  \textnormal{ for some }j\leq n-1\},\quad n\in \N.
 \end{eqnarray*}

Note that, if an arbitrage node of type II is reached, then a strictly positive gain can be achieved in any continuation of the stock price by either investing in the stock ($\varepsilon=1$) or by shortening the stock
($\varepsilon=-1$) for one time period. Similarly, a riskless gain can be achieved at arbitrage nodes of type I, but it can be made strictly positive only along trajectories
in the set $\mathcal{N}$. Finally, up-down nodes have at least one one-period continuation, on which the stock price strictly increases, and one, on which it strictly decreases.

The following proposition explains the behaviour of nodes in the reduced trajectory set after removing  the trajectories in $\mathcal{N}_n$ from $\mathcal{S}$.
\begin{prop}\label{prop:reduction}
  Fix $n\in \mathbb{N}$ and consider 
 \begin{eqnarray*}
 \tau(S)&:=&\inf\{j=1,\ldots, n:\; \mathcal{S}_{(S,j-1)} \textnormal{ is arbitrage node of type I and }  \\ && \quad S_j\neq S_{j-1}  \}\wedge (n+1),\quad S\in \mathcal{S}.
 \end{eqnarray*}
 Then, for every $S\in \mathcal{S}$ and $i=0,\ldots, n$,
 \begin{equation}\label{eq:tau}
  \tau(S)>i \quad \Rightarrow \quad \exists_{\tilde S\in {\mathcal{S}}\setminus \mathcal{N}_n}: \quad (S_0,\ldots,S_i)=(\tilde S_0,\ldots,\tilde S_i).
 \end{equation}
 Moreover, in the reduced trajectory set $\tilde{\mathcal{S}}:=\mathcal{S}\setminus \mathcal{N}_n$ the nodes up to time  $n-1$ can be classified
 as follows: For every
 $i\leq n-1$ and $ S\in \tilde{\mathcal{S}}$, the node $\tilde{\mathcal{S}}_{(S,i)}$ is up-down (resp. flat), if the node ${\mathcal{S}}_{( S,i)}$
 is up-down (resp. flat or arbitrage of type I) in the original trajectory set.
\end{prop}
\begin{proof}
 We prove \eqref{eq:tau} by backward induction on $i=n,\ldots,1$.
If $\tau(S)>n$, then $S\in \tilde{\mathcal{S}}$, and so the statement is trivial for $i=n$. For the induction step from $i+1$ to $i$, assume that
$\tau(S)>i$. If even $\tau(S)>i+1$, the induction hypothesis applies immediately. So we only need to consider the case $\tau(S)=i+1$. Then, there
is an $\bar S\in \mathcal{S}_{(S,i)}$ such that $\bar S_{i+1}=S_i$, because $\mathcal{S}_{(S,i)}$ is an arbitrage node of type I. This implies
$\tau(\bar S)>i+1$ and, by induction hypothesis, there is an $\tilde S\in \tilde{\mathcal{S}}$ such that $(\bar S_0,\ldots,\bar S_{i+1})=(\tilde S_0,\ldots,\tilde S_{i+1})$.
Since the trajectories $S$ and $\bar S$ coincide up to time $i$, the induction step is complete.

We now turn to the classification of the nodes in the reduced trajectory set. If
$S\in \tilde{\mathcal{S}}$, $j\leq n-1$ and $\mathcal{S}_{(S,j)}$ is an up-down node in the original trajectory set, then there are $S^{(1)}, S^{(2)}\in  \mathcal{S}_{(S,j)}$
such that $S^{(1)}_{j+1}>S_j$ and $S^{(2)}_{j+1}<S_j$. Since $\tau(S^{(\iota)})>j+1$ ($\iota=1,2$), Eq. \eqref{eq:tau} implies that we may choose $\tilde S^{(1)}, \tilde S^{(2)}$
from $\tilde{\mathcal{S}}$
which have the same initial segment as $S^{(1)}, S^{(2)}$ up to time $j+1$. Hence, $\tilde{\mathcal{S}}_{(S,j)}$ is also an up-down node in the `reduced' trajectory set.
Similarly one can show that $\tilde{\mathcal{S}}_{(S,j)}$ is a flat node, if $\mathcal{S}_{(S,j)}$ is flat or arbitrage of type I.
\end{proof}

The proofs of the following (technical) lemmas
can be found in the Appendix.

\begin{lem}\label{lem:1}
 If (nL) holds, then the trajectory set $\mathcal{S}$ has no arbitrage nodes of type II and $\mathcal{N}$ is a null set.
\end{lem}

\begin{lem}\label{lem:4}
Fix $n_0\in \mathbb{N}$. If $\mathcal{S}$ has no arbitrage nodes of type II up to time $n_0-1$ and $(V,n,H)$ is a positive simple portfolio with maturity $n\leq n_0$, then
 the value process satisfies $\Pi^{V,n,H}_j(S)\geq 0$ for every $j\in \No$ and $S\in \mathcal{S}$.
\end{lem}

\begin{lem}\label{lem:3}
 If (nL) holds in the trajectory set $\mathcal{S}$, then (nL) holds in each of the shifted conditional spaces $\mathcal{S}^{(S,j)}$.
\end{lem}

The following lemma, which is concerned with the aggregation of  generalized portfolios and  slightly extends Lemma 3 in \cite{ferrando}, plays a crucial role in the proof of Theorem \ref{thm:cond_Koenig}.

\begin{lem}\label{lem:2}
 Fix $n\in \N$. Suppose that $\mathcal{S}$ has no arbitrage nodes of type II up to time $n-1$, and that $(V_m, n_m, H_m)_{m\in \No}$ is a generalized portfolio such that
 $
 V:=\sum_{m=1}^\infty V_m<\infty.
 $
   Then, for every $i\leq n-1$ and $S\in \mathcal{S}\setminus \mathcal{N}_n$, the series
 $\sum_{m=0}^\infty H_{m,i}(S_{i+1}-S_i)$
 converges in the real numbers, and there is 
 a finite sequence of nonanticipating functions $(G_j)_{j\le n-1}$ such that for every $i\leq n-1$ and $S\in \mathcal{S}\setminus \mathcal{N}_n$
 $$
 G_i(S)(S_{i+1}-S_i)= \sum_{m=0}^\infty H_{m,i}(S_{i+1}-S_i).
 $$
\end{lem}

We are now ready to present the proof of Theorem \ref{thm:cond_Koenig}.
\begin{proof}[Proof of Theorem \ref{thm:cond_Koenig}]

 In view of Lemma \ref{lem:3}, it is sufficient to prove (K) in the original trajectory set $\mathcal{S}$.

 Suppose $f:\mathcal{S} \rightarrow [0,+\infty)$ is a financial position with finite maturity $n_f$ and $\bar\sigma(f)<\infty$.  By Proposition \ref{prop:Koenig}
 it is sufficient to show the following property: Whenever $(V_m,n_m,H_m)_{m\in \No}$ is a generalized portfolio such that
 \begin{equation}\label{eq:if}
 \sum_{m=0}^\infty V_m<\infty \quad \textnormal{ and }\quad \sum_{m=0}^\infty \Pi^{V_m,n_m,H_m}_\infty \geq f \textnormal{ on }\mathcal{S},
 \end{equation}
 then there is a sequence $(f_m)_{m\in \N}$ in $\mathcal{E}_+$ such that
 \begin{equation}\label{eq:then}
 \sum_{m=1}^\infty I(f_m)=\sum_{m=0}^\infty V_m  \quad \textnormal{ and }\quad \sum_{m=1}^\infty f_m \geq  f \textnormal{ on }\mathcal{S}.%\sum_{m=0}^\infty \Pi^{(V_m,n_m,H_m)}_\infty 
 \end{equation}

As a first step we shall prove that we may assume without loss of generality that the generalized portfolio in \eqref{eq:if} satisfies $n_m=n_f$
for every $m\in \mathbb{N}_0$. We argue as follows:
  Assume that there is a trajectory $S^*$ such that
 $$
 f(S^*)- \sum_{m=0}^\infty \Pi^{V_m,n_m,H_m}_{n_f}(S^*)>0,
 $$
 where convergence of the series in $(-\infty, +\infty]$ is guaranteed by  Lemmas \ref{lem:1} and \ref{lem:4}. 
 
  Consider the following functions, each of which represents the terminal wealth of a simple portfolios in the shifted conditional space $\mathcal{S}^{(S^*,n_f)}$

 \begin{eqnarray*}
 g_0(S)&=&(-f(S^*)+\Pi^{V_0,n_0,H_0}_{n_f}(S^*)) \\ &&+\sum_{i=0}^{n_0-n_f-1} H_{0,i+n_f}(S^*_0,\ldots,S^*_{n_f-1}, S_0,\ldots, S_i) (S_{i+1} -S_i),
 \\ g_m(S)&=&\Pi^{V_m,n_m,H_m}_{n_f}(S^*)\\ &&+\sum_{i=0}^{n_m-n_f-1} H_{m,i+n_f}(S^*_0,\ldots,S^*_{n_f-1}, S_0,\ldots, S_i) (S_{i+1} -S_i),
 \end{eqnarray*}
 $S\in \mathcal{S}^{(S^*,n_f)}$.
  
 Then, for every  $S\in\mathcal{S}^{(S^*,n_f)}$, $S'=(S^*_0,\ldots ,S^*_{n_f},S_1,,S_2,\ldots)$ belongs to $\mathcal{S}$. Thus,
 \begin{eqnarray*}
  g_m(S)= \Pi^{V_m,n_m,H_m}_{n_f}(S^*)+\sum_{i=n_f}^{n_m-1} H_{m,i}(S') (S'_{i+1} -S'_i)=\Pi^{V_m,n_m,H_m}_\infty(S')\geq 0
 \end{eqnarray*}
for every $m\geq 1$, and, similarly,
$$
g_0(S)=-f(S')+ \Pi^{V_0,n_0,H_0}_\infty(S'),
$$
since $f$ has maturity $n_f$.
Hence, $\sum_{m=0}^\infty g_m$ is the terminal wealth of a generalized portfolio in the shifted conditional space $\mathcal{S}^{(S^*,n_f)}$, which
additionally superhedges 0, because
 $$
 \sum_{m=0}^\infty g_m(S)=-f(S')+\sum_{m=0}^\infty \Pi^{V_m,n_m,H_m}_\infty(S')\geq 0
 $$
 for every $S\in\mathcal{S}^{(S^*,n_f)}$.  Its initial endowment
 $$
 -f(S^*)+\sum_{m=0}^\infty \Pi^{V_m,n_m,H_m}_{n_f}(S^*)
 $$
 is, however, negative. Hence the Leinert condition is violated in the shifted conditional space $\mathcal{S}^{(S^*,n_f)}$ -- a contradiction to (nL).
 By passing from $\Pi^{V_m,n_m,H_m}_\infty$ to $\Pi^{V_m,n_m,H_m}_{n_f}$, if necessary, we may and do assume without loss
 of generality that $n_m=n_f$ for every $m\in \mathbb{N}_0$.

 Now we denote $\tilde{\mathcal{S}}=\mathcal{S}\setminus \mathcal{N}_{n_f}$ and
 consider
 \begin{eqnarray*}
 \tau(S)&:=&\inf\{j=1,\ldots, n_f:\; \mathcal{S}_{(S,j-1)} \textnormal{ is arbitrage node of type I and }  \\ && \quad S_j\neq S_{j-1}  \}\wedge (n_f+1),\quad S\in \mathcal{S}.
 \end{eqnarray*}

 Thanks to Lemmas \ref{lem:2} and \ref{lem:1}, we find functions $g_i:\DR^{i+1}\rightarrow \DR$, $i=0,\ldots, n_{f}-1$, such that
 \begin{equation*}
\sum_{m=0}^\infty \sum_{i=0}^{n_f-1} H_{m,i}(S) (S_{i+1} -S_i)=\sum_{i=0}^{n_f-1} g_i(S_0,\ldots S_i) \;(S_{i+1}-S_i),\quad S\in \tilde{\mathcal{S}}.
 \end{equation*}
  We now consider, for $i=0,\ldots n_f-1$,
  $$
  G_i(S):=\left\{\begin{array}{cl} g_i(S_0,\ldots, S_i),  & \textnormal{if }  \mathcal{S}_{(S,i)} \textnormal{ is up-down and } \tau(S)>i,   
  \\ 0,& \textnormal{otherwise.}
  \end{array}   \right. 
  $$
  Then, the finite sequence $(G_i)_{i=0,\ldots, n_f-1}$ is still nonanticipating and satisfies 
  \begin{equation}\label{eq:0006}
\sum_{m=0}^\infty \sum_{i=0}^{n_f-1} H_{m,i}(S) (S_{i+1} -S_i)=\sum_{i=0}^{n_f-1} G_i(S) \;(S_{i+1}-S_i),\; S\in \tilde{\mathcal{S}}.
 \end{equation}
Define the function
 $$
 f_1(S):=(\sum_{m=0}^\infty V_m)+ \sum_{i=0}^{n_f-1} G_i(S) \;(S_{i+1}-S_i),\quad S\in {\mathcal{S}}.
 $$
  Then,  $f_1\in \mathcal{E}$ and by \eqref{eq:if} and \eqref{eq:0006}
 $$
 0\leq f(S) \leq  f_1(S),\quad S\in \tilde{\mathcal{S}}.
 $$
 We wish to show that $f_1$ is positive, i.e. an element of $\mathcal{E}^+$, for which it remains to show that $f_1\geq 0$ on $\mathcal{N}_{n_f}$.
 For $S^\#\in \mathcal{N}_{n_f}$, the definition of the $G_i$'s entails that
 $$
 f_1(S^\#)=(\sum_{m=0}^\infty V_m)+ \sum_{i=0}^{\tau(S^\#)-2} G_i(S^\#) \;(S^\#_{i+1}-S^\#_i).
 $$
 Hence, by \eqref{eq:tau}, for every $S^\#\in \mathcal{N}_{n_f}$ there is an $j=0,\ldots, n_f-1$ and an $S\in \tilde{\mathcal{S}}$ such that
 $$
 f_1(S^\#)= (\sum_{m=0}^\infty V_m)+ \sum_{i=0}^{j-1} G_i(S) \;(S_{i+1}-S_i)=:\Pi_j(S).
 $$
 $\Pi_j$ can be considered as the wealth process of a simple portfolio with nonnegative terminal wealth $(f_1)_{|\tilde{\mathcal{S}}}$ on the `reduced' trajectory set
 $\tilde{\mathcal{S}}$. Since  $\tilde{\mathcal{S}}$ has no arbitrage nodes of type II up to time $n_f-1$ by Proposition \ref{prop:reduction} and Lemma \ref{lem:1}, $\Pi_j(S)\geq 0$ for every $j=0,\ldots, n_f-1$ and $S\in \tilde{\mathcal{S}}$
 by Lemma \ref{lem:4}.
 This completes the proof that $f_1\in \mathcal{E}^+$.

 To summarize, we have constructed an $f_1\in \mathcal{E}^+$ such that
 \begin{equation}\label{eq:0004}
  f_1\geq f {\bf 1}_{\tilde{\mathcal{S}}},\quad  I(f_1)=\sum_{m=0}^\infty V_m.
 \end{equation}
We next consider the function
$$
f_2(S)= \sum_{i=0}^{n_f-1} |S_{i+1}-S_{i}| \; {\bf 1}_{\{\mathcal{S}_{(S,i)} \textnormal{ is an arbitrage node of type I} \}} .
$$
Note that $f_2\in \mathcal{E}^+$ with $I(f_2)=0$, since it can be realized by starting with zero initial endowment and holding either 1 or -1 shares of the stock for the next
time period, whenever the trajectory reaches an arbitrage node of type I. As $f_2(S^\#)>0$ for every $S^\#\in  \mathcal{N}_{n_f}$, we obtain in view of \eqref{eq:0004}
$$
f_1+\sum_{m=2}^\infty f_2 \geq f \; \textnormal{ on }\mathcal{S},\quad I(f_1)+\sum_{m=2}^\infty I(f_2)=\sum_{m=0}^\infty V_m.
$$
Hence, the sequence $(f_1,f_2,f_2,f_2,\ldots)$ satisfies \eqref{eq:then}.
\end{proof}

\section{Martingale measures and complete trajectory sets}\label{sec:martingale}

In this section, we establish some connections between the classical notion of  existence of a martingale measure and the continuity conditions (L) and (K) of non-lattice integration. We will focus on martingale measures which have support on a tree in a suitable way.

To this end, we denote by $\R^{\No}$  the space of all real valued sequences
indexed by $\No$ and
by $\mathcal{B}(\R^{\No})$ its Borel $\sigma$-field (which is generated by cylinder sets). Writing $\mathfrak{A}=\mathcal{B}(\R^{\No})\cap   \mathcal{S}$ for the trace-$\sigma$-field
of  $\mathcal{B}(\R^{\No})$ on the trajectory set $\mathcal{S}$, we may, thus, consider the measurable space
$
(\mathcal{S}, \mathfrak{A}).
$
We denote the coordinate process by 
$$
T_i:\mathcal{S}\rightarrow \R,\quad (S_j)_{j\in \No} \mapsto S_i,\quad i\in \No,
$$
and note that each $T_i$ is $\mathfrak{A}$-measurable.

\begin{defi}
 We say that the trajectory set $\mathcal{S}$ has a \emph{martingale measure}, if there is a probability measure $Q$ on $(\mathcal{S}, \mathfrak{A})$ such that
 the coordinate process $(T_i)_{i\geq 0}$ (which describes the stock price evolution) is a martingale with respect to its natural filtration $(\mathcal{F}_{i})_{i\in \No}$, i.e. $\mathcal{F}_i:=\sigma(T_0,\ldots, T_i)$ for every $i$.
 The set of all such martingale measures for $\mathcal{S}$ is denoted by $\mathcal{M}$. A martingale measure $Q\in \mathcal{M}$ is said to have \emph{finite support
 at finite maturities}, if there is a set $\mathcal{S}'\in \mathfrak{A}$ such that $Q(\mathcal{S}')=1$ and such that  the set of all nodes
 $\{\mathcal{S}_{(S,j)}:S\in \mathcal{S}'\}$, through which trajectories in $\mathcal{S}'$ pass at time $j$, is a finite set for every $j\in \No$. We write $\mathcal{M}_0$
 for the set of martingale measures with finite support at finite maturities. For any fixed $Q\in \mathcal{M}$, we write $\bar{\mathfrak{A}}_Q$ for the completion of the $\sigma$-field
 $\mathfrak{A}$ by the null sets of $Q$ and $E_Q[f]$ for the expectation with respect to $Q$ of a $\bar{\mathfrak{A}}_Q$-measurable financial position, provided
 it exists in $[-\infty,+\infty]$. Finally,
 \begin{eqnarray*}
 E_Q^*[f]&:=&\inf\{E_Q[h]:\; h \;\bar{\mathfrak{A}}_Q\textnormal{-measurable},\; f\leq h  \textnormal{ on }{\mathcal{S}},\; E_Q[h] \textnormal{ exists in } \\ &&\quad[-\infty,+\infty]\}
 \end{eqnarray*}
 stands for the \emph{outer expectation} of any financial position $f:\mathcal{S} \rightarrow [-\infty,+\infty]$.
\end{defi}

The next theorem entails that the existence of a martingale measure with finite support at finite maturities implies Leinert's condition.
\begin{thm}\label{thm:mart_measure_Leinert}
 Suppose $\mathcal{M}_0\neq \emptyset$. Then (L) holds and
 \begin{equation}\label{eq:supQ_norm}
 \sup_{Q\in \mathcal{M}_0} E_Q^*[|f|]\leq \bar\sigma(f)\leq  \|f\|
 \end{equation}
 for every financial position $f$.

 Moreover, for every  $Q\in \mathcal{M}_0$,
 $\mathcal{L}^1_{(K)}\subset \mathcal{L}^1(\mathcal{S}, \bar{\mathfrak{A}}_Q,Q)$ and
 $$
 \int_{(K)} f= E_Q[f],\quad f\in \mathcal{L}^1_{(K)}.
 $$
\end{thm}
\begin{rem}\label{rem:L_martingale}
The converse of Theorem \ref{thm:mart_measure_Leinert} fails. Leinert's condition (L) 
does not even imply the existence of a martingale measure. Example \ref{exmp:I_neq_sigma} applies as a counterexample. Indeed, we have already shown that (L) holds. Now suppose that a martingale measure $Q$ exists. 
Then the equation $E_Q[{\bf 1}_{\{T_n=2\}}(T_{n+1}-T_n)]=0$ yields $Q(\{S^{(u,n)}\})=0$ for every $n\in\N$, and, thus, $Q(\{\mathcal{S}_d\})=1$. This in turn implies $E_Q[T_1-T_0]<0$ -- a contradiction to the martingale property of $T$.
\end{rem}

\begin{proof}[Proof of Theorem \ref{thm:mart_measure_Leinert}]
 Consider  a simple strategy $(V,n,H)$ and fix a martingale measure $Q\in \mathcal{M}_0$ with finite support at finite maturities. Then, for every $j\in \mathbb{N}_0$ there are trajectories $\tilde S^{1,j},\ldots,\tilde S^{K_j,j}$ such that, for every $j\in \No$, the nodes $\mathcal{S}_{(\tilde S^{1,j},j)},\ldots,
 \mathcal{S}_{(\tilde S^{K_j,j},j)}$ are pairwise disjoint and 
 $$
 Q\left(\bigcup_{k=1}^{K_j} \mathcal{S}_{(\tilde S^{k,j},j)} \right)=1.
 $$
 Consider the functions
 $$
 \tilde H_j(S):=\sum_{k=1}^{K_j} H_j(\tilde S^{k,j}){\bf 1}_{\mathcal{S}_{(\tilde S^{k,j},j)} }(S),\quad j=0,\ldots, n-1.
 $$
 Then, each of the functions $H_j$ is $\mathcal{F}_j$-measurable and bounded. 
 Hence,
 the stochastic process $(\Pi^{V,n,\tilde H}_j)_{j\geq 0}$
 is a martingale with respect
 to $Q$, since it is a martingale difference of a bounded adapted process with respect to the $Q$-martingale $T$. In particular,
 \begin{equation}\label{eq:0000}
 V=E_Q[\Pi^{V,n,\tilde H}_\infty].
 \end{equation}
 We now compare $\Pi^{V,n,\tilde H}_\infty$ to the terminal payoff $\Pi^{V,n, H}_\infty$ of the original simple portfolio $(V,n,H)$. Note that 
 \begin{eqnarray*}
 &&\Pi^{V,n, H}_\infty(S)-\Pi^{V,n,\tilde H}_\infty(S)\\ &=&\sum_{j=0}^{n-1} \left(H_j(S)-\sum_{k=1}^{K_j} H_j(\tilde S^{k,j}){\bf 1}_{\mathcal{S}_{(\tilde S^{k,j},j)} }(S)\right)(S_{j+1}-S_j),
 \end{eqnarray*}
 and, thus, 
 $$
 \{S:\; \Pi^{V,n, H}_\infty(S)\neq \Pi^{V,n,\tilde H}_\infty(S)\}\subset \bigcup_{j=0}^{n-1} \left(\mathcal{S}\setminus \bigcup_{k=1}^{K_j} \mathcal{S}_{(\tilde S^{k,j},j)}\right).
 $$
 The set on the right-hand side is a $Q$-null set in $\mathfrak{A}$, and, thus, 
 the set on the left-hand side is a $Q$-null set in $\bar{\mathfrak{A}}_Q$. Hence, $\Pi^{V,n, H}_\infty
 \in \mathcal{L}^1(\mathcal{S}, \bar{\mathfrak{A}}_Q,Q)$ and $E_Q[\Pi^{V,n, H}_\infty]=E_Q[\Pi^{V,n,\tilde H}_\infty]$. In view of \eqref{eq:0000}, we have shown the following: (LOP) holds, $\mathcal{E}\subset \bigcap_{Q\in \mathcal{M}_0}\mathcal{L}^1(\mathcal{S}, \bar{\mathfrak{A}}_Q,Q)$ and
 \begin{equation}\label{eq:0001}
  E_Q[g]=I(g),\quad g\in \mathcal{E},\; Q\in \mathcal{M}_0.
 \end{equation}
 By \eqref{eq:0001} and the classical Beppo-Levi theorem we obtain for every $Q\in \mathcal{M}_0$ and $f:\mathcal{S}\rightarrow [-\infty,+\infty]$
 \begin{eqnarray*}
  \bar \sigma (f)&=&  \inf\left\{E_Q\left[\sum_{m=0}^\infty f_m\right]:\;f_0\in \mathcal{E},\;f_m\in \mathcal{E}^+\,(m\geq 1),\;
f\leq \sum_{m=1}^\infty f_m \textnormal{ on }\mathcal{S} \right\} \\ &\geq & E_Q^*[f]
 \end{eqnarray*}
 which implies \eqref{eq:supQ_norm} (up to the trivial inequality $\bar \sigma (f)\leq \|f\|$). Choosing $f\equiv 0$, we obtain $\bar\sigma(0)\geq 0$, i.e. Leinert's condition (L) holds.

 Since the inner expectation $-E^*_Q[-f]$ is always dominated by the outer expectation
 $E^*_Q[f]$, we observe that for every $Q\in \mathcal{M}_0$ and $f\in \mathcal{L}_{(K)}^1$
 $$
 E^*_Q[f]\leq \bar\sigma(f) =-\bar\sigma(-f)\leq -E^*_Q[-f]\leq E^*_Q[f].
 $$
  Hence, both the outer and the inner expectation of $f$ under $Q$ coincide with $\int_{(K)} f$. Therefore, $f\in  \mathcal{L}^1(\mathcal{S}, \bar{\mathfrak{A}}_Q,Q)$
  and
 $$
  E_Q[f]=\int_{(K)} f.
 $$
\end{proof}

The next example illustrates that the existence of a martingale measure with finite support at finite maturities does not imply 
K\"onig's condition (K) in general.
\begin{exmp}
 Consider the trajectory set 
 $$
 \mathcal{S}=\{S^{(u,n)}, S^{(d)}, S^{(0)}:\quad n\in \mathbb{N} \}
 $$
 where $S^{(0)}_i=1$ for every $i\geq 0$,
 $$
 S^{(u,n)}_0=1,\quad S^{(u,n)}_1=2,\quad S^{(n)}_i=2 \textnormal{ for $i<n+1$, and } S^{(n)}_i=4 \textnormal{ for $i\geq n+1$},
 $$
 and
 $$
 S^{(d)}_0=1,\quad S^{(d)}_i=0 \textnormal{ for $i\geq 1$}.
 $$
 The up-branch $\mathcal{S}_u=\{S^{(u,n)}:\; n\in \N\}$ behaves as in Example \ref{exmp:I_neq_sigma}, and thus $\|{\bf 1}_{\mathcal{S}_u}\|=0$ by the same argument as there. 
 If the stock price does not increase between time zero and time one, it can either stay constant at 
 its initial value 1 forever (trajectory $S^{(0)}$) or can decrease to the value 0 at time 1 and then stay at that value 
 (trajectory $S^{(d)}$). Clearly the point mass $Q_0$ on $S^{(0)}$ is the unique martingale measure for this trajectory set (noting that uniqueness can be argued as in Remark \ref{rem:L_martingale}), and it has finite support at finite maturities. We are going to show that 
 \begin{equation}\label{eq:0008}
 \bar\sigma({\bf 1}_{\{S^{(d)}\}} ) =0,\quad \bar I({\bf 1}_{\{S^{(d)}\}} ) =\frac{1}{2}.
 \end{equation}
 In view of Proposition \ref{prop:Koenig}, this shows that K\"onig's condition (K) is violated. 
 We now establish the first identity in \eqref{eq:0008}. Since (L) is satisfied by Remark \ref{rem:constant}, we obtain $\bar\sigma({\bf 1}_{\{S^{(d)}\}} )\geq \bar \sigma(0) \geq 0$. Now consider the following terminal payoffs $f_0 \in \mathcal{E}$, $f_n\in \mathcal{E}^+$, $n\geq 1$, of simple portfolios with zero initial endowment, which are given by 
 $$
 f_0(S)=-(S_1-S_0),\quad f_n(S)=\frac{1}{2}(S_{n+1}-S_n),\;n\geq 1
 $$
 and satisfy 
 $$
 \sum_{m=0}^\infty f_m=  {\bf 1}_{\{S^{(d)}\}}.
 $$
 Hence, $\bar \sigma({\bf 1}_{\{S^{(d)}\}}) \leq 0$. For the second identity in \eqref{eq:0008},
 suppose that $(V_m,n_m,H_m)_{m\in\N}$ is a positive generalized portfolio with $V:=\sum_m V_m<\infty$ such that 
 $$
 \sum_{m=1}^\infty \Pi^{V_m,n_m,H_m}_\infty(S)\geq {\bf 1}_{\{S^{(d)}\}}(S),\quad S\in \mathcal{S}.
 $$
 By Lemmas \ref{lem:2} and \ref{lem:4}, there is a constant $G_0\in \mathbb{R}$ such that
 $$
 V+G_0(S_1-S_0) \geq {\bf 1}_{\{S^{(d)}\}}(S),\quad S\in \mathcal{S},
 $$
 since $\Pi^{V_m,n_m,H_m}_\infty(S^{(d)})=\Pi^{V_m,n_m,H_m}_1(S^{(d)})$. Inserting 
 $S=S^{(u,1)}$ and $S=S^{(d)}$  
  in the previous inequality, yields
 \begin{eqnarray*}
  V+G_0\geq 0 \quad \textnormal{and}\quad V-G_0\geq 1.
 \end{eqnarray*}
Hence, $V\geq 1/2$ and, consequently, $\bar I({\bf 1}_{\{S^{(d)}\}}) \geq 1/2$. Finally, the simple portfolio with 
initial endowment $V=1/2$ and maturity $n=1$, which shortens $1/2$ shares of the stock ($H_0=-1/2$), leads to the terminal wealth
$$
f(S)=\frac{1}{2}(1-S_1+S_0)=\left\{ \begin{array}{cl} 1, & S=S^{(d)}, \\ \frac{1}{2}, & S=S^{(0)},
\\ 0, & S\in \mathcal{S}_u, \end{array} \right.
$$
and, thus, superhedges ${\bf 1}_{\{S^{(d)}\}}$. Therefore, $\bar I ({\bf 1}_{\{S^{(d)}\}}) \leq 1/2$, and the proof of \eqref{eq:0008} is complete.

We note in passing, that the simple portfolio with terminal wealth $-(S_1-S_0)$ constitutes 
an arbitrage with respect to the $\|\cdot\|$-null sets in this example, but not with respect to the null sets of the unique equivalent martingale measure $Q_0$. 
\end{exmp}

Despite of this counterexample, we can apply martingale arguments to establish 
K\"onig's continuity condition. The first result in this direction 
is an immediate consequence of Theorems \ref{thm:mart_measure_Leinert} and \ref{thm:cond_Koenig}.
\begin{cor}\label{cor:mart_Koenig}
 If every shifted conditional space $\mathcal{S}^{(S,j)}$ has a martingale measure with finite support at finite maturities, then (nL) and (nK) hold.
\end{cor}

Our final result provides simple-to-check sufficient conditions for the existence of a martingale measure with finite support at finite maturities in every shifted conditional space, and thus for (K).
It builds on the concept of completeness of a trajectory set as introduced in \cite{ferrando}:

Given a sequence $(S^n)_{n\ge 0}$ in $\mathcal{S}$ satisfying
\begin{equation}  \label{splittingSequence}
~~~S^n_i = S^{n+1}_i~~~0 \leq i \leq n, \;n\in \No,~~~~~~~~
\end{equation}
define
$$
\lim_{n\rightarrow \infty} S^n:=(S^i_i)_{i\in \No}.
$$
We say that $\mathcal{S}$ is \emph{complete}, if in such situation the limit $\lim_{n\rightarrow \infty} S^n$ always is a member of $\mathcal{S}.$

\begin{thm}\label{thm:mm}
 Suppose  $\mathcal{S}$ is complete and has no arbitrage nodes of type II.
 Then, $\mathcal{M}_0 \neq \emptyset$, and (nL) and (nK) hold.
\end{thm}

\begin{proof}
 We first show that $\mathcal{M}_0 \neq \emptyset$.
 The proof relies on a binomial tree construction in conjunction with  the  Daniell-Kolmogorov extension theorem.

  Let $P_0$ be the Dirac measure on $S_0$. Suppose $P_{n-1}$ is already constructed as a probability measure on $(\DR^n,\mathcal{B}(\DR^n))$
 such that its support consists of at most $2^{n-1}$ vectors $x\in \DR^n$ which are initial segments of trajectories in $\mathcal{S}$. This means,
 for every $x\in \supp P_{n-1}$ there is an $S\in \mathcal{S}$ such that $(x_1,\ldots, x_n)=(S_0,\ldots, S_{n-1})$. The induction step from time $n-1$ to time $n$
 goes as follows: For any fixed  $x^{(k)}\in \supp P_{n-1}$ choose an $S\in \mathcal{S}$ such that $x^{(k)}=(S_0,\ldots, S_{n-1})$.
 If the node $\mathcal{S}_{(S,n-1)}$ is an arbitrage node of type I or flat, let
 $$
 x^{(k,1)}=(S_0,\ldots, S_{n-1}, S_{n-1})
 $$
and define
$$
p_{k,1}=P_{n-1}(\{x^{(k)}\}),
$$
 i.e. conditionally on reaching the node $\mathcal{S}_{(S,n-1)}$, we choose the constant continuation to time $n$ with probability 1.

 Otherwise (since there are no arbitrage nodes of type II by assumption) we may choose $S^{u}, S^{d}\in \mathcal{S}_{(S,n-1)}$ such that $S^u_n>S_{n-1}$ and
 $S^d_n<S_{n-1}$.  Let
 $$
 x^{(k,1)}=(S_0,\ldots, S_{n-1}, S^u_{n}),\quad x^{(k,2)}=(S_0,\ldots,S_{n-1}, S^d_{n}).
 $$
 We assign the conditional probabilities in the unique way which makes the one-period model starting at $S_{n-1}$ and moving either
 to $S^u_n$ or to $S^d_n$ `fair', i.e. we define $\hat p_{k,\iota}$ via
 \begin{equation}\label{eq:one_period_martingale}
 \hat p_{k,1}+ \hat p_{k,2}=1,\quad (S^u_n-S_{n-1})\hat p_{k,1}+ (S^d_n-S_{n-1})\hat p_{k,2}=0,
 \end{equation}
 and consider
 $$
 p_{k,\iota}=P_{n-1}(\{x^{(k)}\})\hat p_{k,\iota},\quad \iota=1,2.
 $$
 Apparently, the $p_{k,\iota}$'s constitute a probability mass function on the at most $2^n$ vectors $x^{(k,\iota)}$ in $\DR^{n+1}$. Thus, there is a
 unique probability measure $P_n$ on $(\DR^{n+1},\mathcal{B}(\DR^{n+1}))$ satisfying $P_n(\{x^{(k,\iota)}\})=p_{k,\iota}$ for all possible choices of $k,\iota$.
 This finishes the inductive construction.

The above construction enforces that the family of probability measures $(P_n)_{n\geq 0}$ is consistent, i.e.
for every $n\in \mathbb{N}$ and $B\in  \mathcal{B}(\DR^{n})$
$$
P_n(B\times \DR)=P_{n-1}(B).
$$
Hence, by the Daniell-Kolmogorov extension theorem, there is a unique probability measure $P$ on $(\DR^{\No}, \mathcal{B}(\DR^{\No}))$
such that, for every $n\in \mathbb{N}$ and $B\in  \mathcal{B}(\DR^{n})$,
$$
P(\{S\in \DR^{\No}: \; (S_0,\ldots, S_{n-1})\in B\})=P_{n-1}(B).
$$
Denote
$$
{\mathcal{S}'}= \bigcap_{n\in \mathbb{N}} \{S\in \DR^{\No}: \; (S_0,\ldots, S_{n-1}) \in \supp P_{n-1}\}
$$
and observe that ${\mathcal{S}'}$ belongs to $\mathcal{B}(\DR^{\No})$ as a countable intersection of cylinder sets.
Moreover, the probability measure $P$ is supported on ${\mathcal{S}'}$, because, for every $n\in \mathbb{N}$,
\begin{eqnarray*}
&& P(\{S\in \DR^{\No}: \; (S_0,\ldots, S_{n-1}) \in \supp P_{n-1}\})= P_{n-1}(\supp P_{n-1})=1.
\end{eqnarray*}
Next we apply the completeness of the trajectory set $\mathcal{S}$ in order to show that $\mathcal{S}'$ is a subset of $\mathcal{S}$ and, thus, an element
of the trace-$\sigma$-field $\mathfrak{A}$.
Indeed, if $S\in {\mathcal{S}'}$, then for every $n\in \mathbb{N}$, there is an $x^{(n)} \in \supp P_{n-1}$ such that
$(S_0,\ldots, S_{n-1})=x^{(n)}$. However, by the binomial tree construction, vectors in the support of $P_{n-1}$ are always initial segments of trajectories in $\mathcal{S}$.
Hence, for every $n\in \mathbb{N}$, there is an
$S^{(n)}\in \mathcal{S}$ such that $(S_0,\ldots, S_{n-1})=(S^{(n)}_0,\ldots, S^{(n)}_{n-1})$. The sequence $(S^{(n+1)})_{n\in \mathbb{N}_0}$ then satisfies \eqref{splittingSequence}
and, thus, $S=\lim_{n\rightarrow \infty} S^{(n+1)}$ belongs to $\mathcal{S}$ by completeness.

Recall that any set $A\in \mathfrak{A}$ can be written as $\tilde A\cap \mathcal{S}$ for some $\tilde A\in  \mathcal{B}(\DR^{\No})$. Hence, $A\cap \mathcal{S}'= \tilde A \cap \mathcal{S}' \in \mathcal{B}(\DR^{\No})$
and
$$
Q: \mathfrak{A} \rightarrow [0,1],\quad A\mapsto P(A\cap \mathcal{S}')
$$
defines a probability measure which is supported on ${\mathcal{S}'}$, and, thus, finitely supported at finite maturities.
Finally note that the martingale property
of the coordinate process $(T_n)_{n\geq 0}$ under $Q$ with respect to its own natural filtration is a consequence of \eqref{eq:one_period_martingale}. Hence, $Q\in \mathcal{M}_0$.

Obviously, each of the shifted conditional spaces $\mathcal{S}^{(S,j)}$ inherits completeness and absence of arbitrage nodes
of type II from the original trajectory set. Hence, by the first part of the proof, every shifted conditional space $\mathcal{S}^{(S,j)}$ has a martingale measure with finite support at finite maturities. Now Corollary
\ref{cor:mart_Koenig} applies.
\end{proof}

\appendix
\section{Appendix}

We finally prove the auxiliary lemmas of Section \ref{sec:L_vs_K}.

\begin{proof}[Proof of Lemma \ref{lem:1}]
 Suppose on the contrary that $\mathcal{S}_{(S,j)}$ is an arbitrage node of type II. Hence, for some $\varepsilon \in \{-1,1\}$
 $$
 \varepsilon(\bar S_{j+1}-\bar S_j)>0
 $$
 holds for every $\bar S\in \mathcal{S}_{(S,j)}$. In the shifted conditional space $\mathcal{S}^{(S,j)}$ consider
 the simple portfolio $(0,1,\epsilon)$ with strictly positive terminal wealth
 $$
 f_1(\tilde S)=\varepsilon (\tilde S_1- \tilde S_0)>0,\quad \tilde S\in \mathcal{S}^{(S,j)}.
 $$
 Thus the generalized positive portfolio $(0,1,\epsilon)_{m\in \N}$ superhedges $+\infty$ with initial endowment 0 in
 the shifted conditional space $\mathcal{S}^{(S,j)}$. In view of Proposition \ref{prop:arbitrage}, we arrive at a contradiction to (nL).

 We next show that $\mathcal{N}$ is a null set. For every $m\in \mathbb{N}$, the function
 $$
 f_m(S)= \sum_{i=0}^{m-1} |S_i-S_{i-1}| \; {\bf 1}_{\{\mathcal{S}_{(S,i)} \textnormal{ is an arbitrage node of type I} \}}
 $$
 belongs to $\mathcal{E}^+$ with $I(f_m)=0$, since it can be realised by a simple portfolio with initial endowment 0, by buying 1 or -1 shares of the stock
 whenever an arbitrage node of type I is reached up to time $m-1$. If $S\in \mathcal{N}$, then $f_m(S)>0$ for every sufficiently large $m$ (depending on $S$). Hence,
 $$
 \sum_{m=1}^\infty f_m \geq +\infty \cdot {\bf 1}_{\mathcal{N}},
 $$
 implying $\|{\bf 1}_{\mathcal{N}}\|=0$.
\end{proof}

\begin{proof}[Proof of Lemma \ref{lem:4}]
 Suppose $(V,n,H)$ is a positive simple portfolio with maturity $n\leq n_0$, $S\in \mathcal{S}$ and $i< n$. As the node $\mathcal{S}_{(S,i)}$ is up-down or there
 is an $\tilde S\in \mathcal{S}_{(S,i)}$ such that $\tilde S_{i+1}=S_i$, we find some $S^{(i+1)} \in \mathcal{S}_{(S,i)}$ such that
 $$
 H_i(S)(S^{(i+1)}_{i+1} -S^{(i+1)}_i)\leq 0.
 $$
 Proceeding iteratively, there are $S^{(j+1)} \in \mathcal{S}_{(S^{(j)},j)}$ satisfying
 $$
 H_j(S^{(j)})(S^{(j+1)}_{j+1} -S^{(j)}_j)\leq 0,\quad j=i+1,\ldots,n-1.
 $$
 Then,  with $S^{(i)}:=S$, $S^{(n)}\in \mathcal{S}_{(S^{(j)},j)}$ for every $j=i,\ldots,n-1$ and
 $$
 0\leq \Pi^{V,n,H}_\infty(S^{(n)})=\Pi^{V,n,H}_i(S)+\sum_{j=i}^{n-1} H_j(S^{(j)})(S^{(j+1)}_{j+1} -S^{(j)}_j)\leq \Pi^{V,n,H}_i(S).
 $$
\end{proof}
\begin{rem}
 Lemma \ref{lem:4} remains valid, if the absence of type II arbitrage nodes is replaced by the weaker condition that all nodes up to time $n_0-1$
 are zero-neutral in the sense of \cite{ferrando}. One just combines the argument above with Lemma 1 in \cite{ferrando} in the same way as
 in the proof of their Corollary 2.
\end{rem}

\begin{proof}[Proof of Lemma \ref{lem:3}]
Fix $j_0\in \N$ and $S^{(0)}\in \mathcal{S}$ and consider the shifted conditional space $\bar{\mathcal{S}}=\mathcal{S}^{(j_0,S^{(0)})}$. Each shifted conditional space relative to $\bar{\mathcal{S}}$ is of the form
 \begin{eqnarray*}
 \bar{\mathcal{S}}^{(\tilde S,j)}&=&\{(S_{j+i})_{i\in \No};\; S\in \mathcal{S}^{(j_0,S^{(0)})} \textnormal{ and } (S_0,\ldots, S_j)= (\tilde S_0,\ldots, \tilde S_j)\} \\
 &=& \{ (S_{j_0+j+i})_{i\in \No};\;  S\in \mathcal{S} \textnormal{ and } \\ &&\quad (S_0,\ldots, S_{j_0+j}))=(S^{(0)}_0,\ldots, S^{(0)}_{j_0-1}, \tilde S_0,\ldots, \tilde S_j) \}\\
 &=&\mathcal{S}^{(j+j_0,S')},
 \end{eqnarray*}
 where
 $$
 S'=(S^{(0)}_0,\ldots, S^{(0)}_{j_0-1}, \tilde S_0,\tilde S_1,\ldots)\in \mathcal{S}.
 $$
 Since $\mathcal{S}^{(j+j_0,S')}$ satisfies (L) by the nodewise Leinert condition for the original trajectory set, then so does $\bar{\mathcal{S}}^{(\tilde S,j)}$, which shows that
 the nodewise Leinert condition holds in the shifted conditional space $\mathcal{S}^{(j_0,S^{(0)})}$.
\end{proof}

\begin{proof}[Proof of Lemma \ref{lem:2}]
 Suppose $(V_m, n_m, H_m)_{m\in \No}$ is a generalized portfolio such that
 $
 \sum_{m=1}^\infty V_m<\infty,
 $
 and fix $n\in \N$.
 We define
 $$
 G_i(S):=\left\{\begin{array}{cl} \sum_{m=0}^\infty H_{m,i}(S), & \textnormal{if convergent with value in }\R \\ 0, & \textnormal{otherwise,}\end{array} \right.\quad S\in \mathcal{S}.
 $$
 Then, $(G_i)_{0\leq i \leq n-1}$ is nonanticipating and, for $0\leq i \leq n-1$ and $S\in \mathcal{S}$, the identity
 $$
 G_i(S)(S_{i+1}-S_i)= \sum_{m=0}^\infty H_{m,i}(S)(S_{i+1}-S_i)
 $$
 is valid, whenever the series over $(H_{m,i}(S))_{m\in \No}$ is convergent in $\mathbb{R}$ or $S_{i+1}=S_i$. Now, let $S\in \mathcal{S}\setminus \mathcal{N}_n$ and $0\leq i\leq n-1$.
 By assumption, the node
 $\mathcal{S}_{(S,i)}$ cannot be an arbitrage node of type II. If it is an arbitrage node of type I or if it is flat, then $S_{i+1}=S_i$. It is, thus, sufficient to prove the
 following assertion by induction on $i=0,\ldots, n$:
\\[0.1cm]
 For every $S\in \mathcal{S}\setminus \mathcal{N}_n$ and every $j<i$: If $\mathcal{S}_{(S,j)}$ is an up-down node then $\sum_{m=0}^\infty H_{m,j}(S)$ converges in $\mathbb{R}$.
\\[0.1cm]
 For $i=0$, there is nothing to show. For the induction step from $i$ to $i+1$, assume that $S\in \mathcal{S}\setminus \mathcal{N}_n$ and $\mathcal{S}_{(S,i)}$ is an up-down node. Then, for every
 $j<i$,  $\mathcal{S}_{(S,j)}$ is an up-down node or $S_{j+1}=S_j$. Hence, by the induction hypothesis
 $$
  \sum_{m=0}^\infty H_{m,j}(S)(S_{j+1}-S_j)
 $$
 converges in $\R$ for every $j<i$ and, then so does
 $$
 \sum_{m=0}^\infty  \Pi^{V_m,n_m,H_m}_i(S)= \sum_{m=0}^\infty \left(V_m+\sum_{j=0}^{i-1} H_{m,j}(S)(S_{j+1}-S_j) \right),
 $$
 because the series over the initial endowments converges in $\R$ by assumption. By Lemmas \ref{lem:1} and \ref{lem:4},
 $$
  \sum_{m=0}^\infty  \Pi^{V_m,n_m,H_m}_{i+1}(\tilde S) = \sum_{m=0}^\infty \left(V_m+\sum_{j=0}^{i} H_{m,j}(\tilde S)(\tilde S_{j+1}-\tilde S_j) \right)
 $$
 converges in $(-\infty,+\infty]$ for every $\tilde S\in \mathcal{S}$, and therefore the difference of both series
 \begin{equation}\label{eq:0007}
 \sum_{m=0}^\infty  H_{m,i}(S)(\tilde S_{i+1}-\tilde S_i)
 \end{equation}
 converges in $(-\infty,+\infty]$ for every $\tilde S\in \mathcal{S}_{(S,i)}$. Since $\mathcal{S}_{(S,i)}$ is an up-down node  we find
 an $\tilde S \in \mathcal{S}_{(S,i)}$ such that $\tilde S_{i+1}<\tilde S_i$. Dividing by $(\tilde S_{i+1}-\tilde S_i)<0$ in \eqref{eq:0007} yields that
  $
  \sum_{m=0}^\infty H_{m,i}(S)
  $
  converges in $[-\infty,+\infty)$. A `symmetric' argument choosing instead $\tilde S\in \mathcal{S}_{(S,i)}$ such that  $\tilde S_{i+1}>\tilde S_i$ shows that
  $
  \sum_{m=0}^\infty H_{m,i}(S)
  $
  converges in $(-\infty,+\infty]$. Therefore this series converges in $\R$ and the induction step is completed.
\end{proof}


\begin{thebibliography}{99}


\bibitem{acciaio}

B. Acciaio, M. Beiglb\"ock, F. Penkner, and W. Schachermayer (2016) {\it A model-free version of the fundamental theorem of asset pricing and the super-replication theorem}, Math. Finance {\bf 26}, 233--251.

\bibitem{bartl}
D. Bartl, M. Kupper, D.J. Pr\"omel, and L. Tangpi (2019)
{\it Duality for pathwise superhedging in continuous time}, Finance Stoch. {\bf 23}, 697--728.


\bibitem{bartl2}
D. Bartl, M. Kupper, and A. Neufeld (2020) {\it Pathwise superhedging on prediction sets}, Finance Stoch. {\bf 24}, 215--248.

\bibitem{beiglbock}
M. Beiglb\"ock, A.M.G. Cox, M. Huesmann, N. Perkowski,  and D.J. Pr\"omel (2017) {\it Pathwise superreplication via Vovk's outer measure}, Finance Stoch.{\bf 21}, 1141--1166.

\bibitem{bender1} C. Bender (2012) {\it Simple arbitrage}, Ann. Appl. Probab. {\bf 22}, 2067--2085.


\bibitem{bender} C. Bender, S.E. Ferrando and A.L. Gonzalez (2021), \emph{Conditional non-lattice integration, pricing and superhedging}, preprint.


\bibitem{bender2} C. Bender, T. Sottinen, and E. Valkeila (2008) \emph{Pricing by hedging and no-arbitrage beyond semimartingales}, Finance Stoch. {\bf 
12}, 441--468.


\bibitem{biagini} S. Biagini  and R. Cont (2007). \emph{Model-free representation of pricing rules as conditional expectations}. In: Stochastic
Processes and Applications to Mathematical Finance, World Scientific, Hackensack, NJ, pp. 53--66.



\bibitem{burzoni3}
M. Burzoni, M. Frittelli, Z. Hou, M. Maggis, and J. Ob\l{}\'oj (2019)
{\it Pointwise arbitrage theory in discrete time}, Math. Oper. Res. {\bf 44}, 1034--1057.

\bibitem{burzoni}
M. Burzoni, M. Frittelli, and M. Maggis (2016)
{\it Universal arbitrage aggregator in discrete-time markets under uncertainty}, Finance Stoch. {\bf 20}, 1--50.


\bibitem
{burzoni2} M. Burzoni, M. Frittelli,  and M. Maggis (2017) \emph{Model-free superhedging duality},
   Ann. Appl. Probab. {\bf 27} (3), 1452--1477.


\bibitem{cassese} G. Cassese (2008) {\it Asset pricing with no exogeneous probability measure} Math. Finance {\bf 18}, 23--54.    
   
   
\bibitem{cheridito}
P. Cheridito (2003) {\it Arbitrage in fractional Brownian motion models}, Finance Stoch. {\bf 7}, 533--553.
   
\bibitem{coxHouObloj} A.M.G. Cox, Z. Hou, and  J. Ob\l{}\'oj (2016). \emph{Robust pricing and hedging under trading restrictions and the emergence of local martingale models}, Finance and Stochastics, {\bf 20} (3), 669-704.

%\bibitem{cutland}
%N. L. Cutland and A. Roux (2012). \emph{Derivative Pricing in Discrete Time}, Spinger Undergraduate Mathematics Series. Springer, London.

\bibitem{czichowsky}
C. Czichowsky, R. Peyre, W. Schachermayer, and J. Yang (2018) {\it Shadow prices, fractional Brownian motion, and portfolio optimisation under transaction costs}, Finance Stoch. {\bf 22}, 161--180.

\bibitem
{dalang} R.C. Dalang, A. Morton, and W. Willinger (1990)
\emph{Equivalent martingale measures  and no-arbitrage in stochastic securities market models}, Stoch. Stoch. Rep., {\bf 29} 185--201.

\bibitem{davis}
  M.H.A. Davis and D.G. Hobson (2007) {\it The range of traded option prices},  Math. Finance {\bf 17}, 1--14.


\bibitem{delbaen}
F. Delbaen and W. Schachermayer (1994) {\it A general version of the fundamental theorem of asset pricing}, Math. Ann. {\bf 300}, 463--520.

\bibitem{ferrando} S.E. Ferrando and A.L. Gonzalez (2018), \emph{Trajectorial martingale transforms.
Convergence and integration}, New York J. of Math. {\bf 24}, 702--738.


\bibitem{ferrando2} S.E. Ferrando, A.L. Gonzalez, I. L. Degano, and M. Rahsepar (2019) \emph{Trajectorial market models: arbitrage and pricing intervals},
 Rev. Un. Mat. Argentina,
{\bf 60}, 149--185.

%\bibitem{follmer} H. F\"{o}llmer and A. Schied.
%Stochastic Finance: An Introduction in Discrete Time, %4th Edition. De Gruyter 2016.


\bibitem{guasoni}
P. Guasoni (2006) {\it No arbitrage under transaction costs, with fractional Brownian motion and beyond}, Math. Finance {\bf 16}, 569--582.

\bibitem{guasoni2}
P. Guasoni, M. R\'asonyi, and W. Schachermayer (2008) {\it Consistent price systems and face-lifting pricing under transaction costs}, Ann. Appl. Probab. {\bf 18},  491--520.


\bibitem{kassberger} S. Kassberger and T. Liebmann (2016)  \emph{An alternative axiomatic characterisation of pricing operators}, J. Appl. Probab. {\bf 53}, 1257--1264.

\bibitem%
{jarrow} R.A. Jarrow, P. Protter, and H. Sayit (2009) \emph{No arbitrage without semimartingales},
Ann. Appl. Probab. {\textbf 19}, 596--616.

\bibitem{konig} H. K\"{o}nig (1982). \emph{Integraltheorie ohne Verbandspostulat}, Math. Ann. {\bf 258}, 447--458.

\bibitem
{leinert} M. Leinert (1982)  \emph{Daniell-Stone integration without the lattice condition}, Arch. Math. {\bf 38}, 258--265.

\bibitem{perkowski}
N. Perkowski and D.J. Pr\"omel (2016) {\it Pathwise stochastic integrals for model free finance},  Bernoulli {\bf 22}, 2486--
2520.

\bibitem{peyre}
R. Peyre (2017) {\it Fractional Brownian motion satisfies two-way crossing}, {Bernoulli} {\bf 23}, 3571--3597. 

%\bibitem{prokaj}
%V. Prokaj and J. Ruf (2018) \emph{Local martingales in discrete time}, Electron. %Commun. Probab. {\bf 23}, 11 pp.

\bibitem{riedel} F. Riedel (2015) \emph{Financial economics without probabilistic prior assumptions},
Decis. Econ. Finance {\bf 38}, 75-91.

\bibitem{schoenmakers}
J. Schoenmakers and P. Kloeden (1999) {\it Robust option replication for a Black--Scholes model extended with nondeterministic trends}, {J. Appl. Math. Stoch. Anal.} {\bf 12}, 113--120. 


\bibitem
{shafer} G. Shafer and V. Vovk (2019)
{\it Game-Theoretic Probability: Theory and Applications to Prediction, Science, and Finance}, Wiley.

\bibitem{vovk1}
V. Vovk (2012) \emph{Continuous-time trading and the emergence of probability}, Finance Stoch.
{\bf 16}, 561--609

\bibitem
{vovk} V. Vovk (2015), \emph{It\^o calculus without probability in idealized financial markets},
Lith. Math. J. {\bf 55}, 270--290.



\end{thebibliography}
\end{document}